\documentclass[sn-mathphys,Numbered]{sn-jnl}

\usepackage{graphicx}%
\usepackage{multirow}%
\usepackage{amsmath,amssymb,amsfonts}%
\usepackage{amsthm}%
\usepackage{mathrsfs}%
\usepackage[title]{appendix}%
\usepackage{xcolor}%
\usepackage{textcomp}%
\usepackage{manyfoot}%
\usepackage{booktabs}%
\usepackage{algorithm}%
\usepackage{algorithmicx}%
\usepackage{algpseudocode}%
\usepackage{listings}%
\usepackage{subcaption}

\theoremstyle{thmstyleone}%
\theoremstyle{thmstyletwo}%

\theoremstyle{thmstylethree}%

\begin{document}

\title[Article Title]{DEFT: Distributed, Elastic, and Fault-tolerant State Management of Network Functions}


\author[1]{\fnm{Md Mahir} \sur{Shahriyar}}\email{1605024@@ugrad.cse.buet.ac.bd}
\equalcont{Authors have contributed equally to this work.}

\author[1]{\fnm{Gourab} \sur{Saha}}\email{1605053@@ugrad.cse.buet.ac.bd}
\equalcont{Authors have contributed equally to this work.}

\author[1]{\fnm{Bishwajit} \sur{Bhattacharjee}}\email{1605003@ugrad.cse.buet.ac.bd}
\equalcont{Authors have contributed equally to this work.}

\author*[1]{\fnm{Rezwana} \sur{Reaz}}\email{rezwana@teacher.cse.buet.ac.bd}

\affil*[1]{\orgdiv{Department of Computer Science and Engineering}, \orgname{Bangladesh University of Engineering and Technology}, \orgaddress{\street{West Palashi}, \city{Dhaka}, \postcode{1205},  \country{Bangladesh}}}


\abstract{Network function virtualization is the key to developing elastically scalable and fault-tolerant network functions (e.g. load balancer, firewall etc.). By integrating NFV and SDN technologies, it is feasible to dynamically reroute traffic to new network function (NF) instances in the event of an NF failure or overload scenario. The fact that the majority of network functions are stateful makes the task more challenging. 
State migration and state replication are common approaches in achieving elasticity and fault tolerance. The majority of the studies in the literature either emphasize fault tolerance or elastic scalability while designing a state management system for network functions. In this paper, we have designed a complete state management system, called DEFT, that meets both elasticity and fault-tolerance goals. Our system also supports strong consistency on global state updates. While existing designs rely on a central controller or remote central storage to achieve strong consistency on state updates, DEFT utilizes distributed consensus mechanism to achieve the same. We have done a proof of concept implementation of DEFT and extensively tested DEFT under several model  conditions to evaluate its scalability and performance. Our experimental results show that DEFT is scalable and maintains a considerably high throughput throughout. It incurs minimal performance overhead while achieving strong consistency on state updates.}




\maketitle

\section{Introduction}\label{sec1}

Network Function Virtualization (NFV) has replaced the dedicated hardware for implementing network functions  with virtual machines (VMs) that run on physical servers. Significant benefits of NFV include scalability, elasticity, fault-tolerance, and cost-efficiency. 

NFV allows packet processing to be distributed across multiple virtual network function instances. In the rest of this article, an instance of a network function is referred to as an NF. 
A network function is called \emph{stateful} when the processing of a current packet depends on the  packets that have been processed earlier. This dependency is carried out by updating a predefined set of parameters while processing packets, called \emph{network function states} or simply \emph{states}. 
Newer packets are processed based on the current values of the state parameters updated by the older packets.  We can categorize states into the following two types.
\begin{itemize}
    \item Per-flow states: States that are updated by a single flow are called \emph{per-flow states}. These states are also termed as \emph{local states}.
    \item Global states: States that are updated by multiple flows are called \emph{global  states}. For example, the number of flows originating from a particular source IP address at a particular time interval is an example of a global state.
\end{itemize}

A network function can be implemented by more than one NFs distributed across different physical servers. Traffic can be distributed on a per-flow or per-packet basis on NFs. The event of \emph{elastic scaling} of NFs takes place when one or more NFs get overloaded and hence traffic is relocated to one or more new NF instances. Failure of an active NF also requires redistribution of traffic to other (replica) NFs.

State management of network functions includes  state sharing among the NFs, state migration under scaling operation if states are not shared across NFs, and state recovering due to NF failure. Some  pioneering efforts to design NF state management systems are OpenNF~\cite{gember2014opennf}, Split-Merge~\cite{rajagopalan2013split},  StatelessNF~\cite{kablan2017stateless}, S6~\cite{woo2018elastic} and so on. 

OpenNF~\cite{gember2014opennf} supports state-sharing across NFs through a central controller. When an NF updates any state, the central controller is notified, which in turn copies the updated state across all NFs. If states are not shared across NFs, then state migration is necessary during a scaling event. OpenNF provides a loss-free and order-preserving state migration mechanism. Failures of NFs can be handled through replication. However, dependency on the central controller to facilitate state synchronization and state-sharing can  lead to a single point of failure and limit scalability. 

StatelessNF~\cite{kablan2017stateless} and S6~\cite{woo2018elastic} avoid state migration during scaling events. In StatelessNF~\cite{kablan2017stateless}, states are stored in a low-latency and resilient remote server. Since states are accessible by all NFs, no migration is required under normal and scaling events. However, this approach requires remote state access for per-packet processing. Also maintaining a low-latency and resilient remote server at all times is often challenging.
S6~\cite{woo2018elastic} also avoids state migration as states are stored in a distributed object space and accessible to all NFs. In S6~\cite{woo2018elastic}, states are accessible by all NFs but every NF does not possess all states. A state object stored in one NF can be remotely accessed by another NF. However, read-heavy states may be exported to requesting NF. Thus, S6 does not eliminate state migration completely. The current design of S6 does not consider NF failure.

Fault-tolerance of NF failures can be achieved by replicating NF states. Pioneering efforts that deal with NF failure by NF replication are Pico replication~\cite{rajagopalan2013pico}, FTMB~\cite{sherry2015rollback}, REINFORCE~\cite{kulkarni2018reinforce}. In these works, for each NF there exists a replica NF and states are replicated from the active NF to the replica NF. These works do not consider elastic scaling. Since states are not shared across all replicas, explicit state migration is required in the event of elastic scaling and load balancing. 

Unlike most of the previous works that primarily focus either on elasticity or NF failure, we aim to integrate both NF failures and elastic scaling while designing a state management system for stateful network functions. 

\subsection{Motivation}
\begin{itemize}
    \item Distributed State Management under Elasticity: One of the prime benefits of virtualization is that the number of NFs in the system can grow and shrink dynamically during runtime. Since most network functions (i.e., firewall, IDS) are stateful, designing an elastic state management system for virtual network functions is needed. Most existing works focus on techniques to manage NF states (i.e., how the states are shared and migrated during scaling events) that rely on a central controller or remote storage. Therefore, there is a scope for more research on state management systems reliant on distributed mechanisms.
    \item State Management under NF Failure: Existing works that deal with state management of virtual NFs often ignore NF failures or simply refer to NF replication to recover from failure. However, there exists a need for devising detailed mechanisms showing how state replication should work under a chosen consistency model. 
    \item Integration of Elasticity and NF Failure: 
    A state management system that only deals with elastic scaling, requires focusing on efficient state migration during scaling events and may not incur any performance overhead during normal operation due to state replication from an active NF to its replica NF. On the other hand, a system that only deals with fault-tolerance, requires to deal with replication mechanism to meet the consistency requirement of the system. However, it does not require dealing with active flow migration from one NF to a non-replica NF.

Moreover, the objective of elastic scaling is to ensure the optimal usage of resources. However, fault-tolerance requires replication of NF instances and incurs some performance penalty during normal operation. Thus, there exists a need for a complete state management system that meets both elasticity and fault-tolerance goals.
\end{itemize}

\subsection{Design Challenges}
\subsubsection{Elasticity Challenges}
\begin{enumerate}
    \item[E1.] Loss Free State Migration: When an active flow is migrated from a current instance to a new instance, the new instance must have the necessary states to process the  incoming requests. Any in-flight traffic that reaches the source instance after the migration has started needs to be processed.
    \item[E2.] Order Preserving State Migration: Per-flow state updates must be done in the order in which packets are received by the switch despite active flow reallocation. Global state updates must be done in the same order across all NFs despite flow migration. 
\end{enumerate}

\subsubsection{Fault-tolerance Challenges}
\begin{enumerate}
    \item[F1.] Loss Free Failure Recovery: 
    State preservation must be guaranteed in the event of NF instance failure or node failure by performing state replication during normal operation. Any in-flight traffic that reaches failed NF instance needs to be processed.
    \item[F2.] Order Preserving Failure Recovery: 
    Per-flow state updates must be done in the order in which packets are received by the switch despite active flow relocation due to NF failure. Global state updates must be done in the same order across all NFs despite NF failure.
\end{enumerate}

\subsection{Our Contribution}
\label{prop}
In this paper, we have designed a complete NF state management system, called DEFT, that addresses both elastic scaling and NF failures and holds the following properties. 

Before we explain the properties of DEFT, we define the consistency model for our system. 

Strong Consistency: State updates are seen by all NFs in the same global order and the global update order of a state is consistent with the local update order of that of a state with respect to an NF.
\subsubsection*{DEFT Properties} DEFT holds the following properties.
\begin{enumerate}
    \item[P1.] Replicated Global States: Global states are replicated across all NFs in the system.
    \item[P2.] Distributed State Management: The tasks related to state management are performed without involving any central entity. The update order of global states is solely determined by NFs. Moreover, state migration and failure recovery are done peer-to-peer. 
    \item[P3.] Strongly Consistent Global States: Global states are strongly consistent across NFs.
    \item[P4.] Order preserving Per-flow State Update: Per-flow states are updated in the order in which the corresponding packets were received by the system irrespective of
the flow migration between NFs due to scaling events or NF failure.
    \item[P5.] Loss-Free and order-preserving state migration and failure recovery.
    \item[P6.] Loose Flow-Instance Affinity: An instance keeps record of per-flow states while the instance is actively processing the flow. When an instance fails, it only exhibits fail-stop behavior.
\end{enumerate}

The rest of this article is organized as follows. We present prior works in Section~\ref{literature}. The design of DEFT is presented in Section~\ref{design}. Section~\ref{workflow} demonstrates the working principle of DEFT. We present the mechanisms to achieve elasticity and fault-tolerance, respectively, in Section~\ref{elasticity} and Section~\ref{fault}. In Section~\ref{analysis}, we present a systematic analysis of our design. Section~\ref{Implementation} describes the proof-of-concept implementation of our design, and Section~\ref{Experiments} summarizes our experimental study. Finally, we conclude the paper and draw some future directions of work in Section~\ref{conclusion}.

\section{Literature Review}
\label{literature}
We divide the literature into two groups. One group of work deals primarily with elastic scaling, and the other group of work focuses on fault-tolerance of network functions. 
\subsection{Elastic State Management}
State management during scaling events follows two strategies: state migration and migration avoidance.
\subsubsection{State Migration Strategy} The following papers adopt state migration strategy.

OpenNF~\cite{gember2014opennf}: OpenNF is a control framework that provides coordinated control of both the internal NF state and network forwarding state. OpenNF maintains a northbound and a southbound API to provide its state management and  migration functionalities. The southbound API provides a standard NF interface for a controller to manage events and to export or import states across NFs. The northbound API supports move, copy and share operations on NF states. OpenNF’s move operation transfers both the state and input traffic for a set of flows from one NF instance to another during scaling events, whereas copy and share operations are used to replicate states to multiple NF instances. Copy operation clones states from one NF instance to another and can be used when state consistency across NFs is not required or eventual consistency~\cite{bailis2013eventual, burckhardt2014principles,vogels2008eventually} is desired. Share operation is used when strong~\cite{vukolic2016eventually} or strict consistency~\cite{adve1996shared, mosberger1993memory} across NF instances with respect to the state updates is required. OpenNF provides loss-free and order-preserving state migration mechanisms, which are developed around a central controller.  The central controller is responsible to maintain the order of state updates across NF during state sharing. When an NF updates any state, the central controller~\cite{dixit2013towards} is notified, which in turn copies the updated state across all NFs. 

Split/ Merge~\cite{rajagopalan2013split}: Split/ Merge provides a hypervisor level abstraction to virtual middleboxes~ \cite{carpenter2002middleboxes,sherry2012making}. In Split/ Merge, application states are divided into two broader classes: Internal and External. Internal states are basically application logic and are only needed for the execution of a replica. On the contrary, external states are states maintained by multiple replicas. External states are subdivided into partitioned~(a collection of per-flow states) and coherent (global). Partitioned states are updated by a particular flow whereas coherent states can be updated by multiple flows. Each replica NF processes different flows and stores the corresponding state updates. When one flow is moved from one NF to another, the state is migrated and the forwarding rule is updated in an openFlow-based orchestrator. Any in-flight traffic that arrives at the source NF after the migration has stopped is dropped. Coherent state updates are periodically merged, and so remain eventually consistent across replicas. 

\subsubsection{Migration Avoidance Strategy} The following papers adopt the Migration Avoidance strategy.

StatelessNF~\cite{kablan2017stateless}: StatelessNF decouples packet processing from accessing the states, essentially making the NFs stateless~\cite{szalay2019industrial}. Their system has a separate data store layer that holds the states. There is no affinity of traffic to instance. Traffic can be distributed on a per-flow basis or per-packet basis. Since states are accessible by all NFs, no state migration is required under normal and scaling events.  solutions, those come at the cost of significant additional latency due to frequent remote state access. Moreover, the key component of StatelessNF, the data store, acts as a single point of failure. So, it needs to be resilient and have low latency.

S6~\cite{woo2018elastic}: In S6 states are stored in a distributed shared object (DSO) space such that all NFs have access to them. The DSO space is constructed as a two-layer structure: a key layer and a state object layer~\cite{stoica2003chord}. The key owner keeps track of the location of the state object. Any state object stored in one NF can be located and remotely accessed by another NF without broadcasting. S6 does not eliminate state migration completely as key-space and object-space reorganization can be required due to scaling events. State objects reside on NFs based on their state-instance affinity. If affinity changes, then the state is migrated to another NF. S6 has proposed three different optimization techniques for states that have high update frequency from multiple NFs: 1) method call shipping instead of object migration, 2) caching objects locally for NFs that can tolerate stale data, and 3) replicating objects, updating their local copy, and finally merging them to the object owner. This type of state replication across NFs supports eventual consistency~\cite{bailis2013eventual,burckhardt2014principles,vogels2008eventually}.

DEFT follows a state migration strategy inspired by  OpenNF and Split/ Merge. Unlike Split/ Merge, DEFT deals with in-flight packets and supports loss-free state migration. Unlike OpenNF, the state management of DEFT is not dependent on any central controller.

\subsection{Fault Tolerant State Management}

Pico Replication~\cite{rajagopalan2013pico}:   Pico Replication relies on frequent  checkpointing~\cite{plank1994libckpt} for failure recovery as it avoids packet replay. Packets are processed in batches and released after the successful completion of a checkpointing. Backup restores to the last saved checkpoint upon failure of the primary NF. Pico does not handle in-flight packets which may arrive at the primary during checkpointing. 

FTMB~\cite{sherry2015rollback}: FTMB proposes a failure recovery model, that relies on packet replay with infrequent checkpointing~\cite{plank1994libckpt}.  An input   packet log is maintained at a predecessor node so that packets can be replayed during failure recovery. Packets can be released after they are processed without waiting for the checkpointing to be done. Checkpointing is done at an infrequent rate in order to achieve high throughput during normal operation. Snapshots of the system state are stored during checkpointing. During failure recovery, the snapshots are used to bring the system consistent with the latest checkpoint and packets are replayed to bring the system back to the state before the failure has occurred. 

REINFORCE~\cite{kulkarni2018reinforce}: REINFORCE proposed a fault-tolerant system with batch processing similar to Pico Replication ~\cite{rajagopalan2013pico} and infrequent checkpointing with packet-replay similar to FTMB \cite{sherry2015rollback}. Every node has a predecessor node and a backup node. Packets are buffered at the predecessor node and replayed to the backup node in case of failure.  After all the packets of a batch are processed at the primary, only the values of Transmit Timestamp (TxTs) table are transmitted to the backup node and after that, the processed batch is released. The value of TxTs helps the backup node remember which packets are released from the primary so that in case of failure, the backup node does not release duplicate packets. Strict checkpointing ~\cite{plank1994libckpt} is imposed instead of lazy checkpointing when the processed batch contains non-deterministic~\cite{cachin2016non} packets. REINFORCE provides support for both software failures including the failure of an individual NF instance, and hardware failures such as node and link failures, or power outages. 

The fault-tolerance mechanism of DEFT is inspired by REINFORCE. DEFT combines the merits of the two lines of work and supports both elasticity and fault-tolerance.

\section{DEFT Design}
\label{design}
In our system, a node comprises multiple NF instances, and each NF works either as a primary NF or as the backup of a primary NF. We call the later ones secondary NFs. 

\begin{figure}
    \centering
    \includegraphics[width=\linewidth]{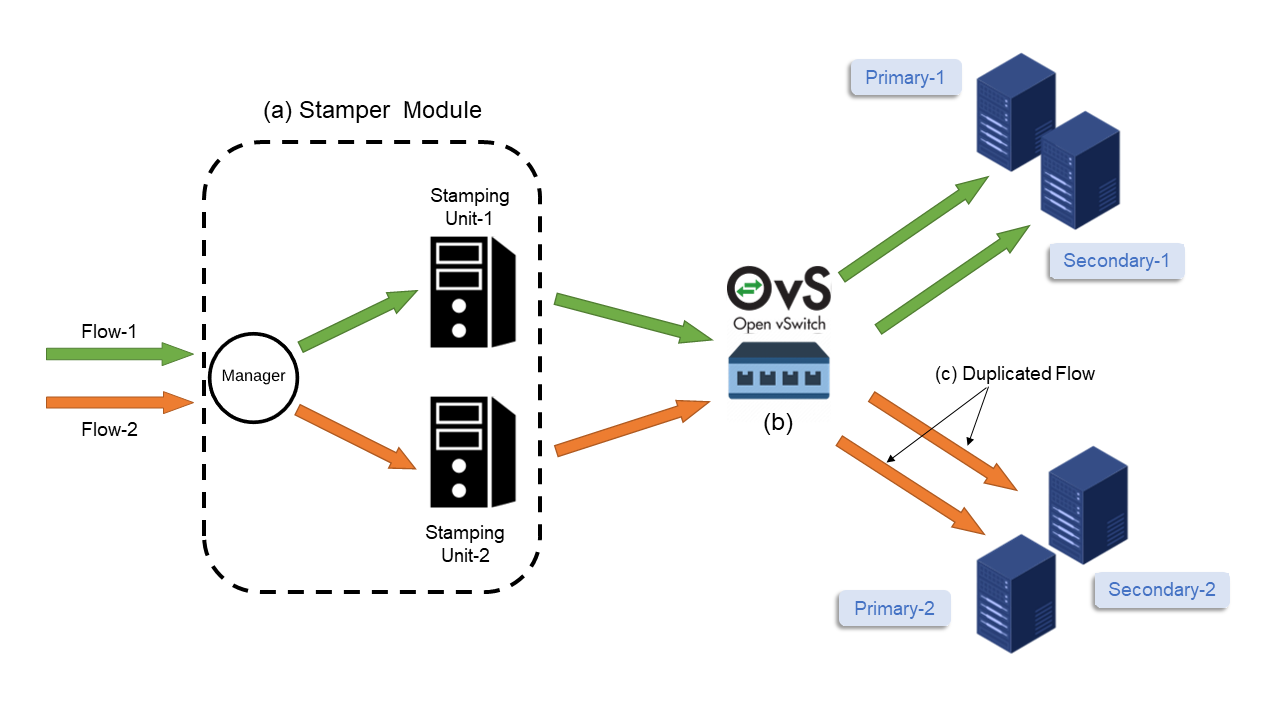}
    \caption{The figure shows a brief overview of DEFT's topology. (a) The stamper module stamps the packets of each flow with a unique \textit{flow ID}. (b) The packets forwarded by the stamper module reach the switch. (c) The switch duplicates each packet and forwards them to the primary NF and the corresponding secondary NF}
    \label{fig:system-topology}
\end{figure}

We show the topology of our system in Figure \ref{fig:system-topology}. 
 
The client's incoming traffic is first routed to the Stamper module. The Stamper module assigns a unique identifier to each incoming packet. Subsequently, the stamped packets are forwarded to the OpenFlow switch, which directs the packets to their intended primary NF. The packets are also duplicated at the switch and sent to the corresponding secondary NFs.

We design our system based on the following assumptions.
\begin{itemize}
    \item An NF instance or a node hosting an NF instance may fail. However, we do not consider any communication link failure.
    \item Our system tolerates fail-stop failure but no byzantine failure.
    \item Packets may get reordered but no packet is lost in the communication channel.
    \item Traffic distribution, load balancing, and detection of NF failures are handled by an SDN controller. The underlying mechanisms to perform these tasks by the SDN controller are out of the scope of this work. Dealing with the failure and recovery of the SDN controller is also out of the scope of this paper. 
    \item We assume per-flow distribution of packets among the primary NFs.
    \item The stamper module is composed of multiple stamping units. The traffic is distributed among these units on a per-flow basis.
    
\end{itemize}
We will discuss more on our failure model in Section~\ref{fault}. 

As part of the SDN controller, there exists a failure detection Unit~(FDU) which is responsible for the detection of any NF failure in the network. The primary function of the FDU  unit is to check the status of all primary NFs and their corresponding secondary NFs by pinging them at regular intervals. Each active instance responds to the ping request. Upon detection of primary NF failure, the FDU unit will inform the corresponding secondary NF to take the role of the primary NF. Since the FDU unit is part of the SDN controller, the failure recovery of this unit is also out of the scope of this work.

\begin{figure}
    \centering
    \includegraphics[width=\textwidth]{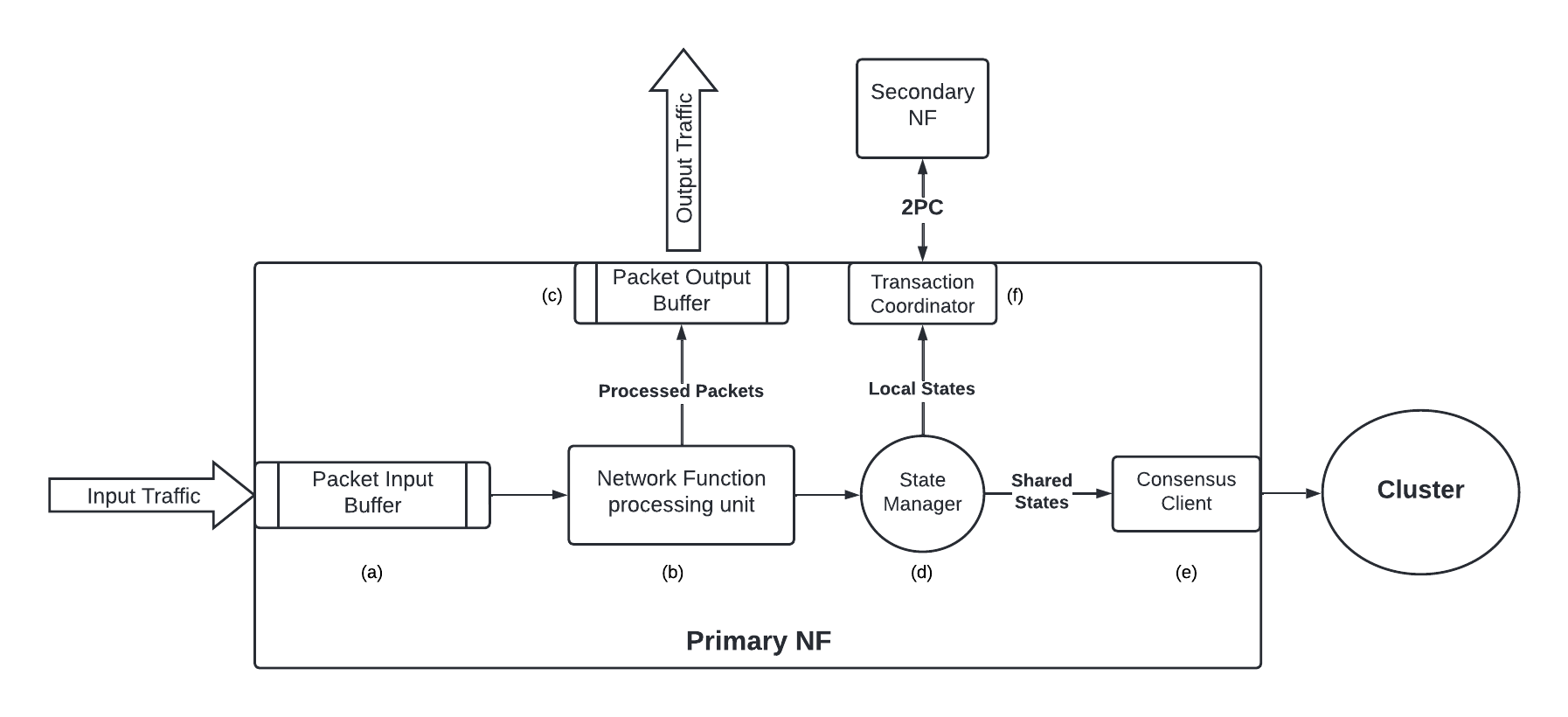}
    \caption{The figure shows the high-level  architecture of an NF. (a) Input traffic reaches the NF and resides inside the input buffer. (b) The processing unit processes each packet. (c) The output buffer stores the processed packets and releases them after each batch. (d) The state manager receives the state information from the processing unit. Then, it forwards the information regarding global states to the consensus client and those regarding local states to the transaction coordinator. (e) The consensus client performs global state updates via consensus. (f) The transaction coordinator performs 2PC after each batch of packets is processed.}
    \label{fig:sys-architecture}
\end{figure}

An NF instance is comprised of the following major components: input buffer, output buffer, network function processing unit, state manager, transaction coordinator, and a client to  an NF cluster. All NFs in the network forms this cluster. 
We present the architecture of an NF instance of  our system in Figure~\ref{fig:sys-architecture}.

An input buffer stores the incoming packets until they are processed and the output buffer holds the processed packets until they are released. We assumed that input buffers are reasonably large. Hence, we do not impose any buffer size restriction while designing DEFT. However, we have shown in Section~\ref{Experiments} the practical size requirement for input buffers with respect to output buffers for varying load.

The network function processing (or simply the processing unit) implements the underlying network function and processes input traffic. The state manager is responsible for storing and sharing states with other NFs in the system. The state manager of a primary NF delegates the responsibility of sharing state updates with the secondary NF to the transaction coordinator. Whenever the primary NF makes the output buffer full, the transaction coordinator initiates the atomic commit between the primary and the secondary NF to share the corresponding state updates. The responsibility of sharing global state updates is delegated to a client to the NF cluster. The cluster is responsible for the management of the global states in our system. Whenever an NF wants to update any global state, it initiates a consensus protocol informing all other NFs within the cluster about the update. 

\section{Workflow of DEFT}
\label{workflow}
In this section, we discuss in detail the working principle of DEFT under normal operations when  scaling or failure events are absent. In our system, packets of the same flow will always be delivered to the same NF whereas different flows will be distributed among multiple NFs. 

\subsection{Packet Stamping}
All incoming packets are stamped at the entry point, which we call the \textit{stamper module}, in our system.
The stamper module consists of multiple stamping units and a stamper manager. The stamper manager is responsible for distributing incoming packets to the appropriate stamping unit based on a predefined hash calculated from the packet's header. This process ensures that packets belonging to the same flow are processed by the same unit. The task of a stamping unit is to assign a unique identifier to the payload of each incoming packet and forward it to the switch. This unique identifier for each flow is called \textit{flow ID}. Each stamping unit assigns \textit{flow ID}-s from a disjoint subset of natural numbers.


\subsection{Packet Duplication} A packet that arrives at the switch from the stamping unit has a unique identifier which is a tuple of two integers (\textit{flow ID}, \textit{per-flow packet counter}). The stamped packet is then forwarded to a primary NF from the switch. The packet is also duplicated and forwarded to the corresponding backup NF.

\subsection{Packet Processing}
We want to achieve order-preserved per-flow state updates and strongly consistent global state updates. To achieve this, we want an NF to process the packets of a particular flow in the order they were received by the stamper module. But a packet might get reordered on the path to NF. The architecture in Figure~\ref{fig:sys-architecture} shows that the packets will reside in an input buffer before entering the processing unit. The input buffer can be considered as a FIFO queue.

The processing unit maintains a HashMap called \texttt{nextExpectedPktID} to select the packet to process from the input buffer. The \texttt{nextExpectedPktID} is a HashMap with \textit{flow ID} as the key and one plus the count of the processed packets so far from the corresponding flow as the value. For example, suppose 5 packets have been processed so far of the flow with \textit{flow ID} = 12. In the HashMap, the entry with key value = 12 will have 6 in its value field. 

A packet with an identifier (\textit{flow ID}, \textit{per-flow packet counter}) \emph{matches}~ \texttt{nextExpectedPktID} if there is an entry in the corresponding HashMap with \textit{flow ID} as its key and  \textit{per-flow packet counter} as its value. For example, a packet with an identifier (11, 3) \emph{matches}~  \texttt{nextExpectedPktID} if there is an entry in the corresponding HashMap with key = 11 and value = 3.

The following steps describe the workflow of the packet processing unit.

\begin{enumerate}
    \item[Step 1:] If there is no space in the output buffer, then no packet is retrieved from the input buffer to process.
    \item[Step 2:] If there exists space in the output buffer, then a packet $p$ is popped from the queue. If the packet identifier of $p$ matches with \texttt{nextExpectedPktID}, then $p$ is processed and placed into the output buffer. Also, the value of the corresponding entry in the \texttt{nextExpectedPktID} is incremented.
    
    Otherwise, $p$ is placed in a HashMap called \texttt{pendingList} with \textit{flow ID} as the key and a priority queue as the value. The priority queue stores the packets that cannot be processed now. The priority queue is sorted by the value field (i.e. \textit{per-flow packet counter}) of the packet identifier. 
    
    Step 1 and Step 2 are repeated. Each time a packet from a particular flow is processed, the \texttt{pendingList} is checked whether any subsequent packet from the same flow is pending to be processed. If the  identifier of a packet $p$ of that flow matches with \texttt{nextExpectedPktID}, then $p$ is processed and removed from the \texttt{pendingList} and \texttt{nextExpectedPktID} is updated accordingly. The process is repeated until there is no pending packet of that flow that can be processed now.
\end{enumerate}
After processing and updating the states of a packet, primary NF will store that packet in the output buffer. The size of the output buffer is denoted as the \textit{batch size}.

\subsection{State Update}
A packet might update both per-flow states and global states in the system. We can also term the per-flow state updates as local updates.

\begin{enumerate}
    \item Global Update: All the global updates need to be shared with the rest of the NFs in the cluster in the same order the primary NF processes them. Whenever a global update is encountered for any packet $p$, the primary initiates the consensus protocol with state update along with the packet identifier of $p$. This identifier will only  be remembered by the corresponding secondary NF. This is required because if the primary fails before completing the current batch, the secondary will reprocess the corresponding batch and update local states skipping the global updates up to the last received packet identifier. A successful return from a consensus protocol marks the end of processing of packet $p$ and the \texttt{nextExpectedPktID} is updated accordingly.  
    
    \item Local Update: DEFT achieves order-preserving local state updates. The primary NF processes the packets of a flow in the same exact order in which the switch receives them. All the packets that leave the processing unit reside in the output buffer and are not immediately released. 
\end{enumerate}

\subsection{State Replication}
When a batch is full, the primary NF shares two pieces of information  with the secondary NF: \texttt{packet clock} and \texttt{state clock}. \texttt{packet clock} consists of  the most recent HashMap \texttt{nextExpectedPktID} after processing a batch and a batch ID $t$, where $t$ is an integer value. For example, after processing the first batch of packets, primary NF will send the current value of the \texttt{nextExpectedPktID} with a batch ID 1 as \texttt{packet clock} to the secondary NF. \texttt{packet clock}  is shared with secondary NF via two-phase commit protocol (2PC)~\cite{2pc}. Packet processing is halted during this time.  Upon committing, the primary NF will release the batch from the output buffer. 

\texttt{state clock} consists of three elements: the most recent HashMap \texttt{nextExpectedPktID}, corresponding state updates, and a batch ID $t$, where $t$ is an integer value. For example, after processing the first batch of packets, primary NF will send the current value of the \texttt{nextExpectedPktID}, corresponding states  with a batch ID 1 as \texttt{state clock} to the secondary NF via 2PC. Packet processing is not halted during this time. Upon committing, the secondary NF will discard the corresponding batch of packets from its input buffer. 

\begin{enumerate}

	\item Reason for sending \texttt{packet clock}: In case of primary NF failure, secondary NF will replay buffered packets to retrieve states. Since \texttt{packet clock} indicates the last batch of packets that the primary NF has released, the secondary NF will not release the same packets. 
	
	\item Reason for sending \texttt{state clock}: When the secondary receives \texttt{state clock} with batch ID $t$,  it can be guaranteed that the last state updates with which it is consistent with the primary NF correspond to the updates by batch $t$ of packets. So, upon failure, the secondary can process packets from the next batch to avoid duplicate updates of the same states.
\end{enumerate}

\section{Elastic Scaling}
\label{elasticity}
When any NF $A$ gets overloaded, some flows need to be directed to a different NF $B$. Only relocation of flows is not enough as NF $B$ does not possess all the states related to these flows. We discuss the scaling procedure below.

We consider that packets of a particular flow will be directed to a particular NF i.e. no packet of the same flow will be distributed among multiple primary NFs. So, if we want to scale, we require some set $S$ of flows that were forwarded towards $A$ to be now directed towards $B$. Let, $s$ be a flow and $s \in S$. We now describe the methodology we follow while migrating flow $s$ from NF $A$ to NF $B$.

First, we start buffering the packets of flow $s$ at NF $A$. Then we migrate the states pertinent to flow $s$ to NF $B$ via 2PC. After the migration is complete, both NF $A$ and $B$ have the necessary states to process the packets of flow $s$. The SDN controller then changes the flow rule at the switch so that the packets of flow $s$ are now forwarded towards NF $B$. So, all new packets of flow $s$ will now arrive at $B$ instead of $A$. Finally, NF $A$ will forward the buffered packets of $s$ towards the switch and these forwarded packets too will be sent towards NF $B$. If there are any in-flight packets of flow $s$, they will be forwarded to $B$ after reaching $A$.

We can show that no packet will be lost during or after state migration and per-flow state updates preserve the order in which the corresponding packets were received by the switch. The first proof is trivial because none of the packets that arrive at $A$ are dropped and directed towards $B$ later on. For the second proof, we have to understand that the new packets of flow $s$ which are sent from the switch to $B$ have higher identifiers than the packets of flow $s$ that are forwarded from $A$ to $B$. So, even though the new packets of flow $s$ sent from the switch to $B$ arrive before the packets forwarded from $A$ to $B$, these packets would reside in the NF $B$'s input buffer. NF $B$ will process a packet of flow $s$ if  the identifier of that packet matches the \texttt{nextExpectedPktID} of NF $B$. Note that during the state migration, NF $A$ will also share its \texttt{nextExpectedPktID} to $B$ and $B$ will update its own \texttt{nextExpectedPktID} accordingly. As a result, NF $B$ knows the count of the last packet of flow $s$ that has been processed by NF $A$.

\section{Fault Tolerance} 
\label{fault}
Achieving fault tolerance is a prime objective of DEFT design. Studies have shown that middlebox-failure \cite{gill2011understanding, potharaju2013demystifying} and software-failure \cite{gunawi2014bugs, gunawi2016does, sahoo2010empirical} happen quite often in a system. In our system, we  consider both NF instance failure (software failure) and node failure.

Before getting into failure recovery we need to clarify some of the assumptions we make in our architecture.

\begin{itemize}
	\item The failure detection unit (FDU) frequently checks for any form of NF failure. Handling the failure of the FDU unit is out of the scope of this work. Thus, NF failure detection is entirely performed by this component. 
	\item Primary NF and the corresponding secondary NF do not reside on the same physical node.
	\item Primary NF and the corresponding secondary NF do not fail simultaneously.
	\item We only consider the crash failure of NFs in our system.
\end{itemize}

Now we discuss our failure recovery mechanism.

\subsection{NF Failure Recovery}
When a primary NF fails, the corresponding secondary NF takes over and continues packet processing. First, the FDU unit acknowledges primary NF $A$ failure and informs the corresponding secondary NF $B$. Next, it assigns NF $B$ as the new primary and NF $C$ as the new secondary NF. Now, NF $B$ has to migrate the state updates along with the last received \texttt{nextExpectedPktID} from NF $A$ and the buffered packets to NF $C$ so that the new secondary NF becomes consistent with the new primary NF. Note that, we would not incur any packet loss here as duplicate packets are always forwarded to the corresponding secondary NF and in this case, all the packets that already reached (or in-flight) to NF $A$ but have not been processed due to the failure of $A$ are also sent to NF $B$. Now, NF $B$ needs to process these buffered packets. The SDN controller changes the flow rule at the switch such that all the packets related to that flow would now be forwarded to NF $B$ and duplicate packets would be sent to the newly assigned secondary NF $C$.

Let the last received \texttt{packet clock} by NF $B$ from NF $A$ has a batch ID $i$. Also, let the last received \texttt{state clock} by NF $B$ from NF $A$ has a batch ID $j$.
Now, one of the following two cases occurs.
 
\begin{enumerate}
	\item Case 1: If $i = j$. In this case, NF $B$ starts processing packets, updating states, and releasing packets according to the \texttt{nextExpectedPktID} value in the \texttt{packet clock}.
	\item Case 2: $i = j + 1$. In this case, NF $B$ starts processing packets and updating states according to the \texttt{nextExpectedPktID} value in the  \texttt{state clock}. However, packets are released according to the \texttt{nextExpectedPktID} value in the \texttt{packet clock}.
\end{enumerate}

All the newly processed packets' per-flow states will be shared with the newly assigned secondary NF $C$ just like before. By following this mechanism, we can both ensure loss-free packet processing and achieve order-preserving per-flow state updates even if primary NF fails at any time.

If any secondary NF fails, the SDN controller would perform the following:

\begin{enumerate}
	\item Assign a new secondary NF and change the forwarding rule at the switch such that duplicate packets will now be forwarded to the newly assigned secondary NF.
	\item Instruct primary NF to share its states with the newly assigned secondary NF via 2PC. Packet processing will be halted during this transaction.
\end{enumerate}
\subsection{Node Failure Recovery}
Our system distributes network function (NF) instances among available nodes to ensure that the primary and secondary instances of the same NF always run on different nodes. This approach guarantees that if a node fails along with all the NFs hosted on it, the system will continue to operate normally. This is because, for every NF impacted by the failed node, there is a backup NF in a different node.

Our system can tolerate simultaneous failure of multiple primary NFs residing in the same node or across several nodes. It can also tolerate simultaneous failure of multiple nodes as long as the failed nodes do not host the primary and backup instances of the same NF.

\subsection{Failure of Stamper Module}
The stamper module consists of multiple stamping units and a stamper manager. 
In the event of a single stamping unit failure, only the flows assigned to that particular unit will be impacted (incoming packets will be dropped). Suppose, the failure of a stamping unit $su$ impacts flow $f$. Upon recovery of $su$, the 
remaining packets of flow $f$ will reach $su$ due to predefined hashing. Flow $f$ will now be recognized as a new flow by $su$.

To minimize the potential impact of a single stamping unit failure, the system can be designed with a larger number of stamping units such that flows are evenly distributed across several units. To minimize the impact of a single node failure, stamping units are distributed across several nodes. Moreover, we can replicate each stamping unit to lower the impact of failure. 

If the stamper manager fails, incoming packets will be dropped momentarily. Upon recovery, the manager ensures that packets belonging to the same flow are always forwarded to the same stamping unit. A backup manager can reduce the impact of failure.

\section{Analysis of the Designed System}
\label{analysis}
In this section, we show that our system holds the properties listed in Section~\ref{prop}. 

\subsection{Analysis of Property P1} 
\textit{Replicated Global States: Global states are replicated across all NFs in the system.}\\
Each NF holds a copy of each global state. When a primary NF $N$ processes a packet $p$ that leads to an update to a global state $GS$, $N$ initiates a consensus protocol that runs among all NFs in the system. Upon successful return from a consensus call, each NF updates the corresponding global state $GS$.
\subsection{Analysis of Property P2}
\textit{Distributed State Management: The tasks related to state management are performed without involving any central entity. The update order of global states is solely determined by NFs. Moreover, state migration and failure recovery are done peer-to-peer}.\\

There mainly three major tasks related to state management; (1) global state update - global states are updated by running consensus among the NFs;
(ii) state migration - the process of state migration is peer-to-peer during a scaling event while the SDN controller is responsible to initiate the scaling event; and (iii) state recovery after an NF failure  is executed by the replica NF by packet reprocessing while the SDN  controller is responsible for failure detection and flow redirection.

\subsection{Analysis of Property P3} 
\textit{Strongly Consistent Global States: Global states are strongly consistent across NFs.}\\

We need to show that (i) global state updates are seen by all NFs in the same global order and (ii) the global update order of a global state $s$ is consistent with the local update order of $s$ with respect to an NF. 

Whenever an NF processes a packet that attempts to update a global state $s$, a consensus protocol is initiated. Upon successful return from the consensus protocol, the corresponding update is seen by all NFs. To prove the second part of strong consistency, we assume that $p$ and $q$ are two packets from two different flows that are being processed by $NF1$. Both $p$ and $q$ updates global state $s$. Packet $p$ is ready to be processed before packet $q$ in $NF1$. When $p$ is being processed, $NF1$ cannot start processing $q$ by the design. Hence $NF1$ initiates a consensus for $q$ only after a successful return from a consensus initiated for $p$. Therefore, in the global update order for state $s$, packet $p$ updates $s$ before packet $q$. This order is also consistent with the local update order of state $s$ with respect to $NF1$.

\subsection{Analysis of Property P4} 
\textit{Order preserving Per-flow State Update: Per-flow states are updated in the order in which the corresponding packets were received by the  system irrespective of
the flow migration between NFs due to scaling events or NF failure.}\\

Assume that two packets $p$ and $q$ belong to the same flow $f$, and $p$ is received before $q$ by the stamper module, which is the entry point of our system. Flow $f$ is send to $NF1$ by switch $sw$. Packets may get reordered from the stamper module to $sw$ or from $sw$ to $NF1$. Without loss of generality, we consider the following three cases. 

\emph{Case 1: $p$ and $q$ are reordered from $sw$ to $NF1$, i.e., $q$ reaches $NF1$ before $p$.}  Since  $p$ is stamped with a lower identifier than $q$ by $sw$ and $NF1$ processes packets according to \texttt{nextExpectedPktID}, packet $p$ will be processed before packet $q$ by $NF1$. 

\emph{Case 2: Migration of flow $f$ from $NF1$ to $NF2$ starts after $sw$ sends $p$ to $NF1$}. Upon receiving $p$, $NF1$ will buffer $p$. In the meantime, suppose packet $q$ is sent to $NF2$ by $sw$. However, $NF2$ cannot process $q$ as $q$ does not match its \texttt{nextExpectedPktID}. After state migration, $NF2$ receives the buffered packets (including $p$) of flow $f$ from $NF1$ and processes them. Since $q$ has a higher identifier than the buffered packets, $q$ will be processed after $p$ in $NF2$.

\emph{Case 3: $NF1$ fails after $sw$ sends $p$ to $NF1$.} Note that packet $p$ will also be sent to the secondary NF of $NF1$ by $sw$. Upon failure of $NF1$, $sw$ will relocate flow $f$ to the secondary NF of $NF1$. The secondary NF of $NF1$ will process packets of flow $f$ in the order in which the packets were stamped. 
\subsection{Analysis of Property P5}
\textit{Loss-Free and order-preserving state migration and
failure recovery.}\\

We need to prove that the designed system guarantees (i) loss-free and order-preserving state migration during scaling events and (ii) loss-free and order-preserving state recovery in the event of NF failure.

Proof of (i): For simplicity, we consider only one flow is relocated and corresponding states are migrated to a destination NF during a scaling event. The source NF buffers the incoming packets as soon as the state migration starts. After the completion of the state migration, the buffered packets are also transferred to the destination NF. Any in-flight packets to the source NF during this transfer, are also sent immediately to the destination NF. Hence, no state is lost as the destination NF has the state updates that occurred before the migration started and all the packets that have been received by the source or destination NF afterward. Moreover, we have already proved in the Property $P4$ that the packets of a flow are processed in the order they were received by the system despite flow migration.

Proof of (ii): For simplicity, we consider only one flow is affected due to an NF failure. In our design, packets are sent from a switch to both primary and secondary NF. After a primary NF processes a batch of packets, the corresponding state updates are sent to the secondary NF. The secondary NF stores the packets until it receives the corresponding state updates. When a primary NF fails, the secondary NF reprocesses the packets for which it has not received any state update resulting in loss-free state recovery. Moreover, we have already proved in the Property $P4$ that the packets of a flow are processed in the order they were received by the system despite NF failure.

\subsection{Analysis of Property P6}
\textit{Loose Flow-Instance Affinity: An instance keeps
record of per-flow states while the instance is actively
processing the flow. When an instance fails, it only
exhibits fail-stop behavior.}\\

An NF instance stores per-flow states in its local memory while it is actively processing the flow. When a primary NF fails while processing any flow $f$, the corresponding secondary NF becomes the new primary. The processing of flow $f$ in the new primary does not depend on the recovery of the old primary. So when an NF instance fails in any node, it stops permanently. New NF instances are required whenever needed based on the traffic load.

\section{Implementation}\label{Implementation}
We have done a proof of concept implementation of our design in a dockerized environment. The network between the stamper module and NFs was set up using the Open vSwitch.

\subsection{Stamper}
The stamper module serves as the main entry point for all incoming traffic in our system, comprising multiple stamping units that operate in parallel to process packets. The module is responsible for attaching a unique identifier to each packet flowing through it. Each flow in our system is identified by a 5-tuple value: \textit{source-ip, destination-ip, source-port, destination-port, and protocol}. Our load-balancing strategy ensures that packets belonging to the same flow are always processed by the same stamping unit. Each stamping unit maintains a mapping between the flow and the number of packets of that flow it has processed so far. Upon receiving a packet, it increments the number of packets that have been stamped of that flow, stamps the new packet with that number, and forwards it to the switch where it is duplicated and sent to a primary and its corresponding secondary NF.\\

For example, if the total number of packets of flow \textit{s} passed through the stamper module is \textit{n}, then the next packet of flow \textit{s} would be stamped with the number \textit{n + 1}. This is necessary for consistency reasons. When an NF receives a packet of a particular flow, it checks whether its \texttt{nextExpectedPktID} for that flow is equal to the packet’s \textit{per-flow packet counter}. If not, then the particular packet has arrived out of order and cannot be processed at the time. To achieve order-preserving per-flow state updates, and to remove central dependency, the stamper module plays a vital role in our design.

\subsection{Network Functions}
Different network functions have different local and global state access and update patterns. So, we built a prototype of a generic network function with controllable parameters in order to test various configurations found in a wide variety of NFs. The controllable parameters are batch size and global state update rate. The batch size denotes how frequently we communicate with the backups. The global state update rate controls how frequently within a batch of packets global state updates (hence consensus) are necessary.\\

Each NF has two threads; the first receives a packet and puts it in the input buffer, and the second takes packets from the input buffer and puts them into the output buffer after processing.

\subsection{Primary and Backup}
For each NF, there is one primary and one backup. The primary and backup are hosted in separate hosts to be resilient against single-host failures. The primary contains the connection information (IP, port) for its backup. We implemented the communication between the primary and backup using RPCs. The backup provides services of \texttt{packet clock} and  \texttt{state clock} update as RPC method. Whenever a primary completes processing a batch, it must atomically commit its current \texttt{packet clock}. We achieve this by initiating the 2PC protocol by the primary. When there is only one backup to the primary, the 2PC protocol simplifies down to just waiting for that only backup to commit synchronously. The primary may need to try (with exponential back-off) several times to successfully replicate \texttt{packet clock} values if the backup is unavailable for some moments.

\subsection{Buffers}
There are two buffers in each NF; input buffer and output buffer. We used a queue as our buffer because we wanted to process the packets and send them in FIFO order. We used python's Queue~\cite{pyqueue} module from the queue class. 

The maximum allowed sizes of both buffers are controllable parameters in our system.

\subsection{Global Update Unit}
We used Hazelcast~\cite{hazeclast} for our consensus module. Hazelcast is an open-source in-memory Data Grid that is peer-to-peer, simple, scalable, fast, and redundant. It uses Raft consensus protocol to provide linearizability even in failure situations. \\

The network functions (NFs) within our system operate together as a cluster, with no master instance to serve as a single point of failure. The modular design of our system allows for simple addition and removal of instances from the cluster as needed. When a new instance is added to the cluster, it automatically discovers the existing cluster members. \\

Our system leverages the Hazelcast Python client to enable network functions (NFs) to connect to the cluster. This client library offers a rich set of distributed data structures, including distributed dictionaries (maps), available in both synchronous and asynchronous versions. To ensure the correctness of our system, we cannot progress further until the consensus is complete and thus used the synchronous version of the distributed dictionary. Moreover, we can turn on the ‘near-cache’ feature on the client at the risk of reading stale data. This feature will ensure that the client has a copy of the updates in its local memory for faster future reads. \\

We used a distributed key-value to store our global states. Whenever an NF needs to update some global states, the python client modifies the corresponding key-value, and initiates the consensus protocol. When the protocol completes, all the NFs in the cluster will get the update and apply them in the same order. 

\subsection{Packet Generation}
We used the Packet Sender~\cite{packet-sender} tool to generate traffic. The tool has a wide range of features for generating TCP and UDP packets. In our current implementation, we focused on UDP traffic and used python's Twisted framework~\cite{twisted} to work as the server for our NFs and stamper module.

\section{Experiments and Results}\label{Experiments}

\subsection{Experimental Setup}

All of our experiments were conducted on an Intel® Core™ i7-10700K CPU @ 3.80GHz CPU with 16 cores and 64GB RAM. The operating system was Ubuntu 18.04.6 LTS. 

\subsection{Evaluation Metrics}
In our experiments, we used various performance metrics. In this section, we will define and describe these metrics.

\subsubsection{Latency}
Latency is defined as the total duration required for a packet to travel from the stamper module to the point at which it is released from the processing NF. This time includes the time spent on the path from the stamper module to the NF, input buffer waiting time before processing, NF processing time, and waiting time in the output buffer before releasing. Let, the time spent on the path from the stamper module to the NF be  $\tau_{path}$, the waiting time in the input buffer before processing be $\tau_{pre}$, processing time be $\tau_{proc}$, and waiting time in the output buffer after processing be $\tau_{post}$. Average latency, $latency$ can be represented by 

\begin{align} \label{eq:latency}
    latency = \frac{\sum_{i=1}^{n}(\tau_{path_i} + \tau_{pre_i} + \tau_{proc_i} + \tau_{post_i})}{n}
\end{align}

where, $n \in \mathbb{N}$ is the number of packets. We track various statistics of the first $n$ packets. For our experiments, we keep track of the statistics of the first 50000 packets irrespective of their final status. Hence, $n$ $=$ $50000$ for all of our experiments is in the following sections.

\subsubsection{Throughput}
Throughput is the total number of packets released per second by an NF. We measure it in \textit{kpps}. Let the start and end times of flows be $\tau_{start}$ and $\tau_{end}$, and the number of packets that exited the NF successfully of flow $j$ be $s_j$ and $|F|$ indicates the total number of flows. Total throughput $throughput_{N}$ can be expressed by equation-\ref{eqn_tpt_n}. 

\begin{align}
    \begin{aligned}
    throughput_{N} = \sum_{j=1}^{|F|}\frac{s_j}{\tau_{end_j} - \tau_{start_j}}
    \label{eqn_tpt_n}
    \end{aligned}
\end{align}


\subsubsection{Packets Dropped}

In our experiments, we keep count of the events when the incoming packets cannot be buffered into an input buffer. This event would not occur if the buffer size is $\infty$. But when buffer size is limited, a packet rate faster than the processing rate may lead to packet drop at the input buffer. Our system does not incur any packet drop at the output buffer as we release the packets immediately when the output buffer is full. Thus, we only quantify the number of packets dropped at the input buffers in the calculation of the total packet drop for a single run of an experiment. 

\subsection{What do we want to evaluate?}
Throughout the evaluation process of DEFT, we want to answer the following key questions. 

\begin{enumerate}
    \item Can we achieve a high packet processing rate despite sharing local states with the replica NF after each batch?
    \item 
    What is the maximum packet loss that DEFT incurs while operating at an optimized packet sending rate? Furthermore, what impact does this level of packet loss have on the system's overall throughput?
    \item How is the latency of the whole system affected due to the local state sharing?
    \item Do the stamper units cause any bottlenecks to the system's overall packet processing?
    \item To what extent do global state updates impact DEFT's ability to handle a high packet rate, and what would be the consequences if a particular packet type requires a significant global update?
\end{enumerate}

\subsection{Local State Update}

For local state updates, our goal is to achieve the lowest magnitude of latency while keeping higher throughput. In the following experiments,  we investigate the optimal values for batch size and buffer size in our system. In successive experiments, we also show the impact of varying the number of incoming flows and stamping units.

\subsubsection{Small vs large batch size. Which one and why?}\label{bs}

As our system processes packets in batches, we first aim to determine the optimal batch size for our system operation. We use Equation \ref{eq:latency} to calculate latency i.e. the average time a packet spends inside DEFT.

\begin{figure} [h!]
  \includegraphics[width=\linewidth]{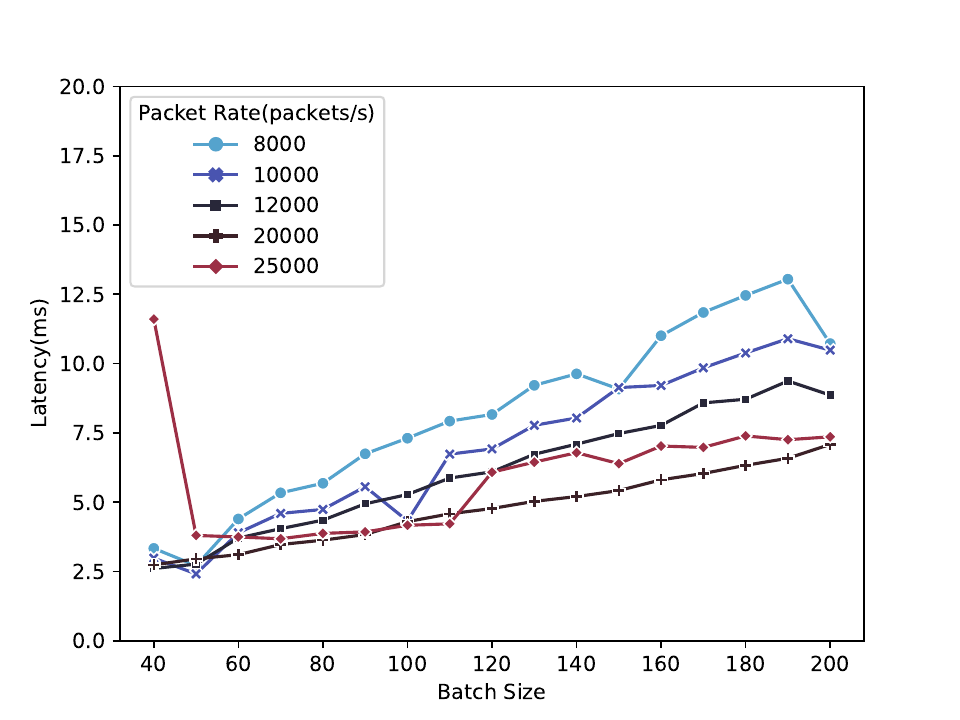}
   \caption{The figure shows the effect of varying batch size and packet rate on DEFT's latency. Increasing the batch size until a threshold reduces latency. After which latency starts to rise again.}
  \label{fig:bs_vs_pr_latency}
\end{figure}

We start with a batch size of 10 packets and send the traffic at 10,000 packets/second to the system. This results in an average latency of 576.11 ms after processing 50,000 packets. As we gradually increase the batch size, we see a sharp drop in latency to 2.41 ms when the batch size gets to 50 packets. After this, the latency does not seem to decrease anymore and gradually starts to increase. At a batch size of 100 packets, we find the system’s latency to be 4.33 ms. Increasing the batch size furthermore results in a linear increase in the latency as well. This is because initially when the batch size is low, the NF spends more time sharing local state updates with the replica than processing packets. But after a certain batch size limit, DEFT’s latency starts to increase as packets now have to stay in the buffer for a longer period of time. Hence, we determine the optimal batch size for DEFT is 50 at which point the system achieves its lowest latency of 2.41 ms. An illustration of this behavior can be seen in Figure \ref{fig:bs_vs_pr_latency}.

\begin{figure}[h!] 
  \includegraphics[width=\linewidth]{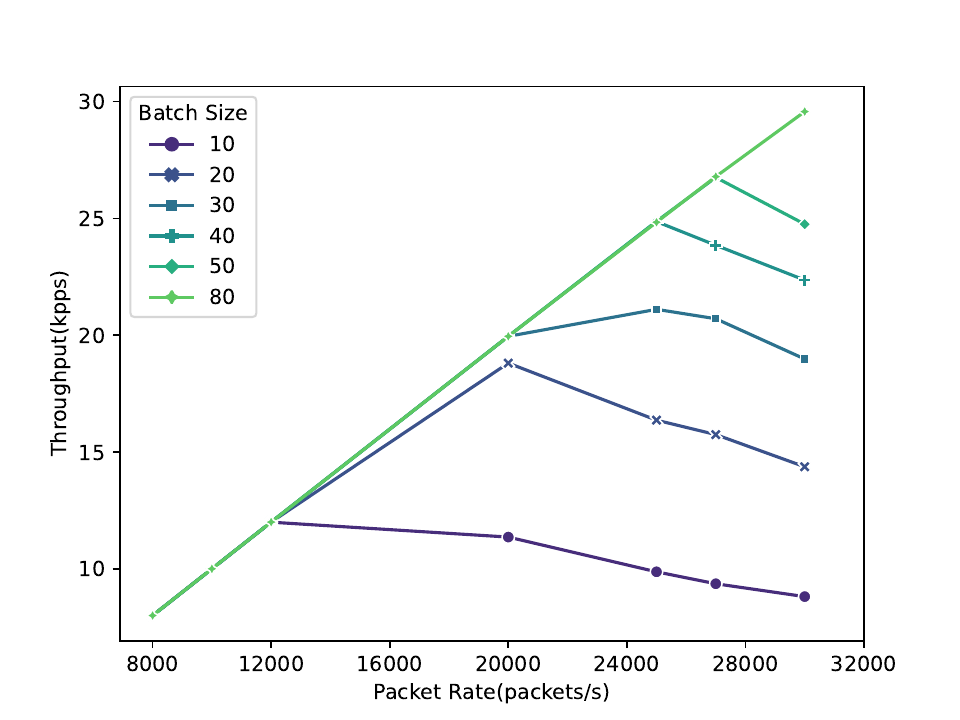}
  \caption{The figure shows the effect of varying batch size and packet rate on throughput. For each batch size, throughput increases linearly until a threshold.}
  \label{fig:bs_vs_pr_throughput_kpps}
\end{figure}

We then try to find the effect of packet rate on this optimum batch size. When traffic is sent at a rate of 8,000 packets/second, DEFT processes 7,950 packets per second. Increasing this traffic to 10,000 packets/second also increases the rate at which DEFT processes them to 9,906 packets per second. This trend in the increase of throughput remains till the packet rate exceeds 27,000 packets/second, beyond which the packet processing rate does not seem to increase linearly as before. Figure \ref{fig:bs_vs_pr_throughput_kpps} illustrates this behavior graphically.

\subsubsection{Buffer size is theoretically infinite. But do we really need it?}
We next evaluate the significance of input buffer size and how it impacts packet drop in DEFT. In theory, a buffer size of $\infty$ will not encounter any form of packet drop. However, we want to find a practical buffer size for our system, which will avoid packet loss.

\begin{figure} 
  \includegraphics[width=\linewidth]{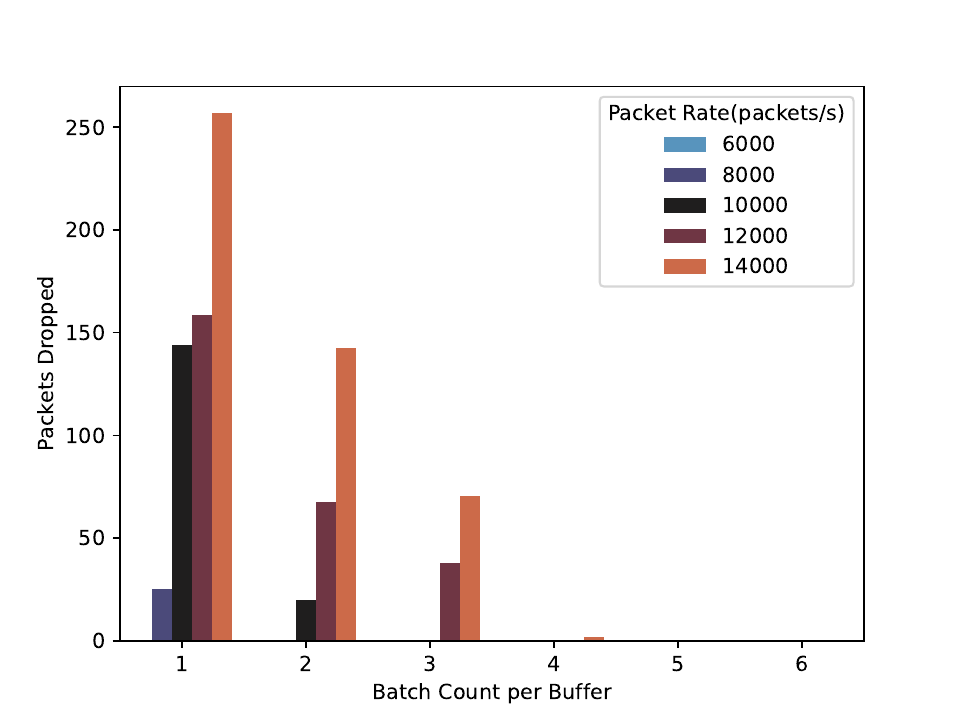}
  \caption{The bar plot shows how changing the buffer size impacts the number of packets dropped while entering the input buffer. When the number of batches in the buffer is greater than 4, no packets are dropped by DEFT.}
  \label{fig:bfs_vs_pr_drop}
\end{figure}

In our system, we measure buffer size as the number of batches of packets it can hold. We start with a buffer size of 1 batch of packets per buffer where each batch comprises 50 packets. When we send the traffic at a rate of 6,000 packets/second, we find that DEFT does not exhibit any form of packet drop. But when we gradually increase the packet rate to 14,000 packets/second, the system drops 257 packets out of the 50,000 packets it processed (about 0.51\%). We then increase the buffer size to 2 batches per buffer keeping the same packet rate and that reduces the number of dropped packets to around 142 (about 0.23\%). Finally, at a buffer size of 5 batches, DEFT manages to process the traffic without dropping any packets. Increasing the buffer size, even more, does not carry any additional significance as the number of packets dropped still remains zero. This concludes that 5 batches of packets per buffer are sufficient for the system to encounter zero packet loss. Figure \ref{fig:bfs_vs_pr_drop} illustrates this behavior.

From Figure \ref{fig:bfs_vs_pr_drop}, we can also see that with an increase in buffer size for any given packet rate, the number of packets dropped starts to reduce. This is because more packets can be accommodated in the buffer. Additionally, for any given buffer size, with an increase in the packet rate, the number of packets dropped starts to increase.

From our experiments, we observe that even for a reasonably high packet rate (e.g. 14,000 pps), a buffer size that is 4 times greater than our optimal batch size is good enough to bring down the packet drop to almost 0.

\subsubsection{Is DEFT actually invariant of flow count?}

We next check to see if our system gets affected by the factors like the number of flows or the number of stamping units. To test this, we set up an environment with 20 NFs and 100 flows. We set up the stamper manager in a way such that it distributes the traffic on a per flow basis evenly among the stamping units. 

We first test to see whether a varying number of stamping units impact DEFT in any way. We start with a single stamping unit and gradually increase the number of stamping units to 20 keeping the packet rate at 12,000 packets/second. With a single stamping unit, the system demonstrates a latency of 3.7 ms after processing 50,000 packets. When we scale up to 6 stamping units, the system shows an increase in its latency to 17.72 ms. Scaling up to 10 stamping units exhibits a similar linear increase in DEFT’s latency to 26.20 ms. Therefore, a trade-off should be made between the degree of fault tolerance and tolerable latency for choosing the number of stamping units in the system.

\begin{figure} 
  \includegraphics[width=\linewidth]{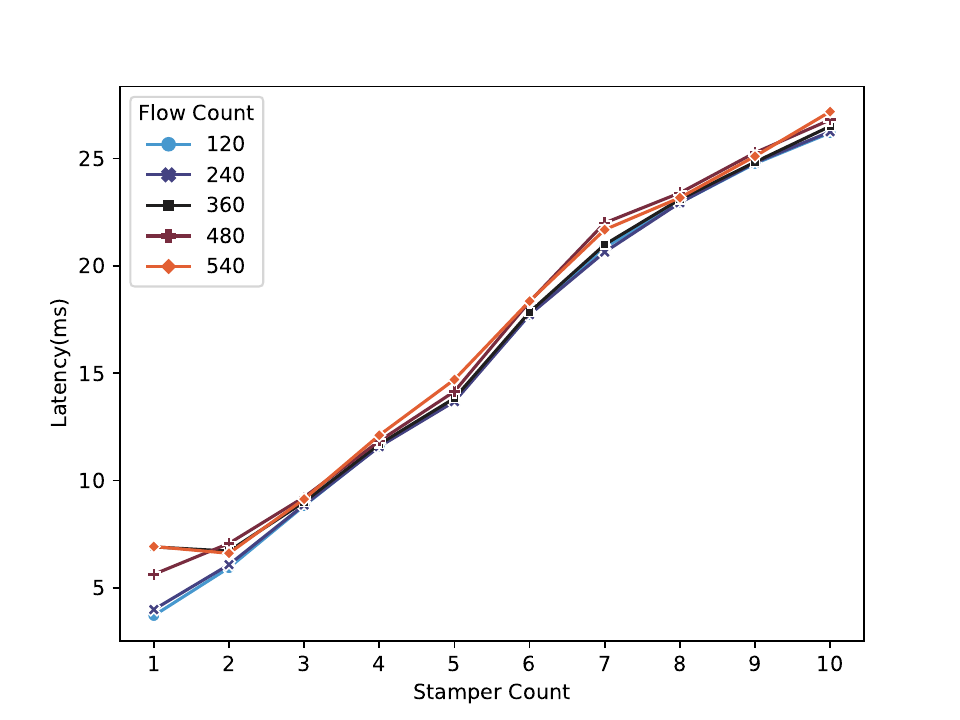}
  \caption{The plot shows the impact of varying stamper and flow count on latency. With an increase in stamper count, latency rises. However, changes to the number of flows do not impact the overall latency. }
  \label{fig:sc_vs_fc_latency}
\end{figure}

However,  the effect of increased stamping units has very little impact on DEFT’s overall throughput. With a single stamping unit, DEFT processes 10,086 packets per second whereas scaling to 10 stamping units resulted in almost the same throughput (10,080 packets per second). More details can be seen in Figure \ref{fig:sc_vs_fc_latency} where we illustrate this behavior graphically.

Next, we test to see whether a varying number of flows impacts DEFT in any way. While keeping the overall system’s traffic rate at 12,000 packets/second we vary the flow count from 120 to 540 but the system’s overall throughput always remains around the value of 10,088. This confirms that DEFT demonstrates resilience to fluctuations in the number of incoming flows and its behavior changes only when the system’s overall traffic rate changes.

\subsection{Global State Update}

Global state updates are costly, as it involves running a consensus algorithm. However, in this section, we showcase DEFT's robustness in handling global state updates and illustrate how the system performs under the burden of running consensus algorithms of greater magnitude.

\subsubsection{How frequently should we perform global updates?}

\begin{figure} 
  \includegraphics[width=\linewidth]{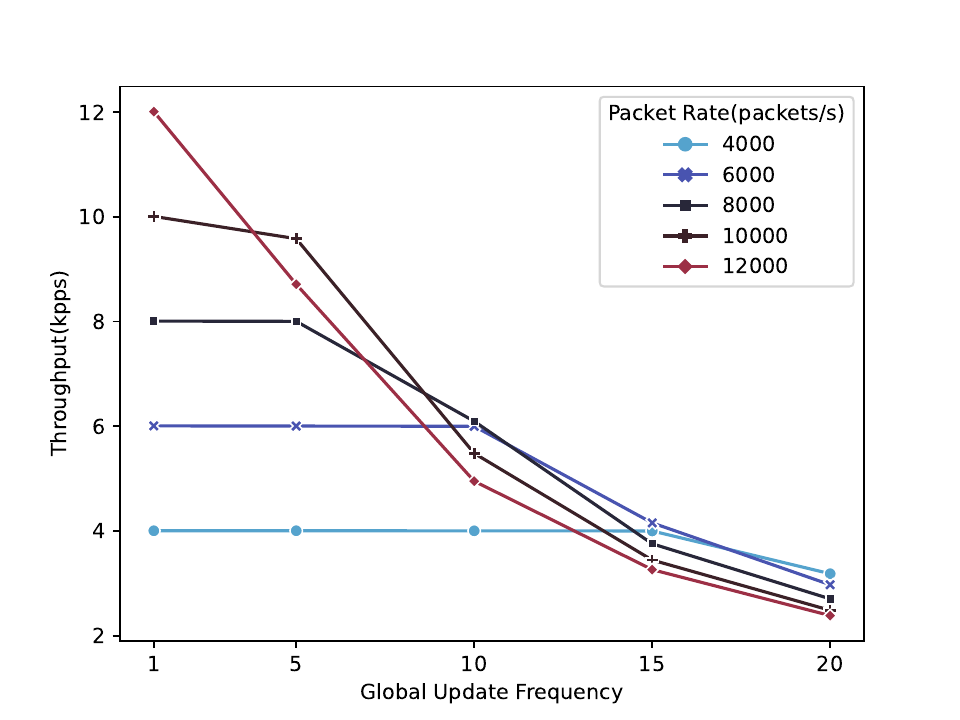}
  \caption{The figure shows the impact of global update frequency on throughput. For a moderate packet rate, the number of global updates has a lesser impact. However, this changes when we increase the packet rate.  }
  \label{fig:uf_vs_pr_throughput}
\end{figure}

We now move our focus on consensus and how it affects the overall system. We start with sending 4,000 packets/second and running a single consensus per batch. The system does not seem to get impacted by this at all and processes the traffic at around 4,000 packets/second. Even if we gradually increase the frequency of global updates per batch, DEFT perfectly processes the traffic at this rate. The same behavior is repeated if we tune up the traffic to 6,000 packets/second. A significant change in throughput is seen when the traffic is raised to 10,000 packets/second with a global update frequency of 10 per batch. Only then, DEFT exhibits a lower packet processing rate of 5,484 packets per second. Increasing the global update frequency to 15 times per batch further decreases the system’s throughput by 37.16\% (3,446 packets/second). Similar behavior is found when the packet rate is increased to 12,000 packets/second. Figure \ref{fig:uf_vs_pr_throughput} illustrates this behavior in detail.

\subsubsection{How well does DEFT tackle a heavy consensus?}

\begin{figure*}[h!]
  \centering
  \begin{subfigure}[t]{\textwidth}
    \centering
    \includegraphics[width=\textwidth]{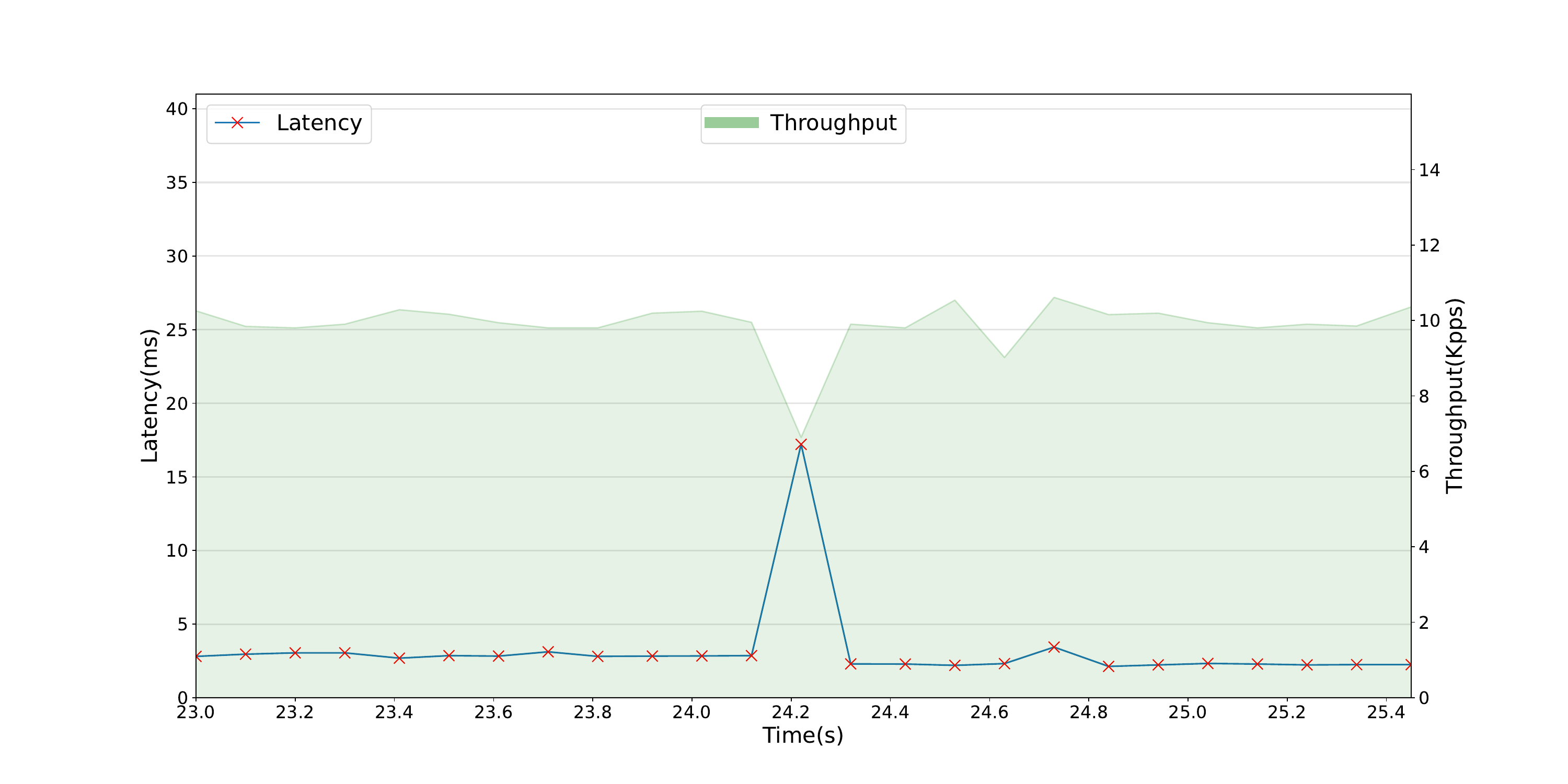}
    \caption{50 Global Updates}
    \label{fig:timelapse_new}
  \end{subfigure}
  
  \begin{subfigure}[t]{\textwidth}
    \centering
    \includegraphics[width=\textwidth]{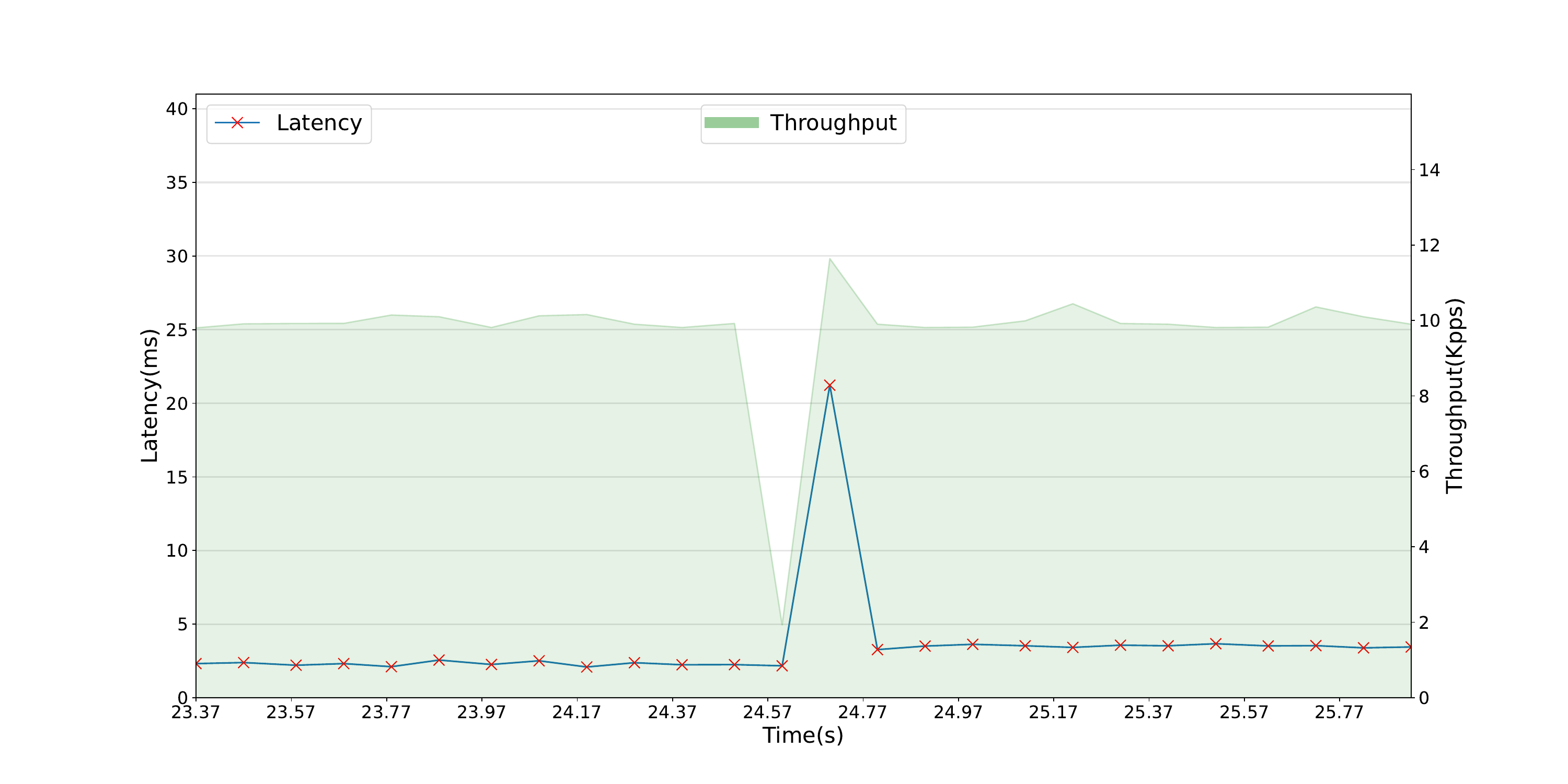}
    \caption{100 Global Updates}
    \label{fig:timelapse_old}
  \end{subfigure}
  
  \caption{The figure shows the impact of a heavy global update on DEFT in a granular timescale. (a) The upper figure demonstrates DEFT's behavior in a state containing 50 global updates. (b) The lower figure demonstrates DEFT's behavior in a state containing 100 global updates. With a heavier global update, latency increase and throughput reduces. }
  \label{fig:timelapse_comparison}
\end{figure*}

Finally, we exhibit DEFT’s behavior at a micro-level and check how resilient DEFT is in the face of a packet containing heavy global state updates. We first send the traffic to an NF at 10,000 packets/second. When 60,000 packets are processed, we issue a global state update 50 times consecutively. Before issuing the global state update, DEFT’s normal latency is around 2.7 ms, but when the global state update takes place 50 times consecutively, the latency rises to 17.21 ms. The reason behind this rise is that the system spends more time in consensus than processing packets during this time. The system’s throughput initially falls for the first 100 ms to 6.9 kpps. After that, when the global update is complete the system returns to its normal form of 10 kpps within the next 100 ms.

We stress test DEFT’s capability further by issuing 100 global state updates consecutively. This time the system exhibits a sharp fall in its throughput for the next 100 ms to 1.9 kpps. When the global state update is complete, the system’s throughput sharply rises and gets back to its normal form of 10 kpps within the next 100 ms. After that, DEFT continues to process packets just like before maintaining a latency of 3.5 ms. This micro-level behavior showcases that even though DEFT faces a temporary rise in latency and drop in throughput during a heavy global state update, the system quickly recovers from this state and resumes normal packet processing. We illustrate this behavior through a timelapse graph in Figure \ref{fig:timelapse_comparison}.

\subsection{Summary of Our Experiments}

To summarize the evaluation section, we tested DEFT in scenarios like amping up the packet sending rate, increasing the rate of global updates, and imposing heavy global updates to stress test the system. From these experiments, we conclude that when packets are sent at a rate of 10,000 packets/second under normal operating conditions DEFT achieves a latency of 2.41 ms with a maximum throughput of 9.9 kpps. DEFT is capable of achieving even higher throughput if the packet rate is increased up to the mark of 27,000 packets/second. Even in heavy global update scenarios, DEFT has shown its capability to retain its throughput instantaneously after the global update finishes.

We also found the following optimal values for the system under the current experimental setup.

\begin{itemize}

\item We found that a batch size of 50 is optimal for DEFT. Decreasing the batch size from this increases the rate of local state updates and increasing it makes the packets wait in the buffer for a longer period of time.

\item The optimal input buffer size for DEFT is 5 batches of packets. Keeping the buffer size to any lower value incurred packet loss in the system.

\item For a moderate packet rate (4,000-6,000 packets/second), DEFT’s throughput and latency are invariant of global state update frequency.

\end{itemize}

\section{Conclusion and Future Work}
\label{conclusion}

While the relevant research on state management systems focuses on either fault-tolerance or elastic scaling, we have designed a complete state management system, DEFT, that is fault-tolerant and supports elastic scaling. DEFT achieves strong consistency on global state updates in a distributed manner. DEFT also guarantees loss-free and order-preserving state migration and failure recovery. Our experiments show that DEFT achieves considerably high throughput under several model conditions. We also observe that DEFT can achieve strong consistency with minimal performance overhead.

In this work, our primary objective was to bring down the design goals under one roof when dealing with both elastic scaling and fault-tolerance and present mechanisms to deal with states without  any central dependency in 
 a comprehensive manner. However, we understand that there are still ways to improve, which we leave as future work. Here are some of the ways.

\begin{enumerate}
    \item We want to integrate practical and commercially available NFs like  Bro IDS~\cite{bro}, and PRADS~\cite{prads} in our implementation and perform experiments to test our system against other state-of-the-art systems and compare the performance.

    \item There is a scope for achieving higher throughput if packet processing inside the NF is parallelized. Therefore, one of the future directions of our system is to incorporate per-flow multithreading at packet processing.
 
    \item Our current configuration has global states shared among all the participants. However, all global states may not be associated with every participant. As such, it adds unnecessary overhead by having more participants than required. We can resolve this issue by having a cluster-based state update for the global states. In this cluster-based design, only the NFs concerned with a  given state will be included in the cluster for that state.
\end{enumerate}

\bibliography{DEFT}


\begin{thebibliography}{32}
\ifx \bisbn   \undefined \def \bisbn  #1{ISBN #1}\fi
\ifx \binits  \undefined \def \binits#1{#1}\fi
\ifx \bauthor  \undefined \def \bauthor#1{#1}\fi
\ifx \batitle  \undefined \def \batitle#1{#1}\fi
\ifx \bjtitle  \undefined \def \bjtitle#1{#1}\fi
\ifx \bvolume  \undefined \def \bvolume#1{\textbf{#1}}\fi
\ifx \byear  \undefined \def \byear#1{#1}\fi
\ifx \bissue  \undefined \def \bissue#1{#1}\fi
\ifx \bfpage  \undefined \def \bfpage#1{#1}\fi
\ifx \blpage  \undefined \def \blpage #1{#1}\fi
\ifx \burl  \undefined \def \burl#1{\textsf{#1}}\fi
\ifx \doiurl  \undefined \def \doiurl#1{\url{https://doi.org/#1}}\fi
\ifx \betal  \undefined \def \betal{\textit{et al.}}\fi
\ifx \binstitute  \undefined \def \binstitute#1{#1}\fi
\ifx \binstitutionaled  \undefined \def \binstitutionaled#1{#1}\fi
\ifx \bctitle  \undefined \def \bctitle#1{#1}\fi
\ifx \beditor  \undefined \def \beditor#1{#1}\fi
\ifx \bpublisher  \undefined \def \bpublisher#1{#1}\fi
\ifx \bbtitle  \undefined \def \bbtitle#1{#1}\fi
\ifx \bedition  \undefined \def \bedition#1{#1}\fi
\ifx \bseriesno  \undefined \def \bseriesno#1{#1}\fi
\ifx \blocation  \undefined \def \blocation#1{#1}\fi
\ifx \bsertitle  \undefined \def \bsertitle#1{#1}\fi
\ifx \bsnm \undefined \def \bsnm#1{#1}\fi
\ifx \bsuffix \undefined \def \bsuffix#1{#1}\fi
\ifx \bparticle \undefined \def \bparticle#1{#1}\fi
\ifx \barticle \undefined \def \barticle#1{#1}\fi
\bibcommenthead
\ifx \bconfdate \undefined \def \bconfdate #1{#1}\fi
\ifx \botherref \undefined \def \botherref #1{#1}\fi
\ifx \url \undefined \def \url#1{\textsf{#1}}\fi
\ifx \bchapter \undefined \def \bchapter#1{#1}\fi
\ifx \bbook \undefined \def \bbook#1{#1}\fi
\ifx \bcomment \undefined \def \bcomment#1{#1}\fi
\ifx \oauthor \undefined \def \oauthor#1{#1}\fi
\ifx \citeauthoryear \undefined \def \citeauthoryear#1{#1}\fi
\ifx \endbibitem  \undefined \def \endbibitem {}\fi
\ifx \bconflocation  \undefined \def \bconflocation#1{#1}\fi
\ifx \arxivurl  \undefined \def \arxivurl#1{\textsf{#1}}\fi
\csname PreBibitemsHook\endcsname

\bibitem[\protect\citeauthoryear{Gember-Jacobson et~al.}{2014}]{gember2014opennf}
\begin{barticle}
\bauthor{\bsnm{Gember-Jacobson}, \binits{A.}},
\bauthor{\bsnm{Viswanathan}, \binits{R.}},
\bauthor{\bsnm{Prakash}, \binits{C.}},
\bauthor{\bsnm{Grandl}, \binits{R.}},
\bauthor{\bsnm{Khalid}, \binits{J.}},
\bauthor{\bsnm{Das}, \binits{S.}},
\bauthor{\bsnm{Akella}, \binits{A.}}:
\batitle{Opennf: Enabling innovation in network function control}.
\bjtitle{ACM SIGCOMM Computer Communication Review}
\bvolume{44}(\bissue{4}),
\bfpage{163}--\blpage{174}
(\byear{2014})
\end{barticle}
\endbibitem

\bibitem[\protect\citeauthoryear{Rajagopalan et~al.}{2013}]{rajagopalan2013split}
\begin{bchapter}
\bauthor{\bsnm{Rajagopalan}, \binits{S.}},
\bauthor{\bsnm{Williams}, \binits{D.}},
\bauthor{\bsnm{Jamjoom}, \binits{H.}},
\bauthor{\bsnm{Warfield}, \binits{A.}}:
\bctitle{Split/merge: System support for elastic execution in virtual middleboxes}.
In: \bbtitle{Proceedings of the 10th USENIX Symposium on Networked Systems Design and Implementation (NSDI 13)},
pp. \bfpage{227}--\blpage{240}
(\byear{2013})
\end{bchapter}
\endbibitem

\bibitem[\protect\citeauthoryear{Kablan et~al.}{2017}]{kablan2017stateless}
\begin{bchapter}
\bauthor{\bsnm{Kablan}, \binits{M.}},
\bauthor{\bsnm{Alsudais}, \binits{A.}},
\bauthor{\bsnm{Keller}, \binits{E.}},
\bauthor{\bsnm{Le}, \binits{F.}}:
\bctitle{Stateless network functions: Breaking the tight coupling of state and processing}.
In: \bbtitle{Proccedings of the 14th USENIX Symposium on Networked Systems Design and Implementation (NSDI 17)},
pp. \bfpage{97}--\blpage{112}
(\byear{2017})
\end{bchapter}
\endbibitem

\bibitem[\protect\citeauthoryear{Woo et~al.}{2018}]{woo2018elastic}
\begin{bchapter}
\bauthor{\bsnm{Woo}, \binits{S.}},
\bauthor{\bsnm{Sherry}, \binits{J.}},
\bauthor{\bsnm{Han}, \binits{S.}},
\bauthor{\bsnm{Moon}, \binits{S.}},
\bauthor{\bsnm{Ratnasamy}, \binits{S.}},
\bauthor{\bsnm{Shenker}, \binits{S.}}:
\bctitle{Elastic scaling of stateful network functions}.
In: \bbtitle{Proceedings of the 15th USENIX Symposium on Networked Systems Design and Implementation (NSDI 18)},
pp. \bfpage{299}--\blpage{312}
(\byear{2018})
\end{bchapter}
\endbibitem

\bibitem[\protect\citeauthoryear{Rajagopalan et~al.}{2013}]{rajagopalan2013pico}
\begin{bchapter}
\bauthor{\bsnm{Rajagopalan}, \binits{S.}},
\bauthor{\bsnm{Williams}, \binits{D.}},
\bauthor{\bsnm{Jamjoom}, \binits{H.}}:
\bctitle{Pico replication: A high availability framework for middleboxes}.
In: \bbtitle{Proceedings of the 4th Annual Symposium on Cloud Computing},
pp. \bfpage{1}--\blpage{15}
(\byear{2013})
\end{bchapter}
\endbibitem

\bibitem[\protect\citeauthoryear{Sherry et~al.}{2015}]{sherry2015rollback}
\begin{bchapter}
\bauthor{\bsnm{Sherry}, \binits{J.}},
\bauthor{\bsnm{Gao}, \binits{P.X.}},
\bauthor{\bsnm{Basu}, \binits{S.}},
\bauthor{\bsnm{Panda}, \binits{A.}},
\bauthor{\bsnm{Krishnamurthy}, \binits{A.}},
\bauthor{\bsnm{Maciocco}, \binits{C.}},
\bauthor{\bsnm{Manesh}, \binits{M.}},
\bauthor{\bsnm{Martins}, \binits{J.}},
\bauthor{\bsnm{Ratnasamy}, \binits{S.}},
\bauthor{\bsnm{Rizzo}, \binits{L.}}, \betal:
\bctitle{Rollback-recovery for middleboxes}.
In: \bbtitle{Proceedings of the 2015 ACM Conference on Special Interest Group on Data Communication},
pp. \bfpage{227}--\blpage{240}
(\byear{2015})
\end{bchapter}
\endbibitem

\bibitem[\protect\citeauthoryear{Kulkarni et~al.}{2018}]{kulkarni2018reinforce}
\begin{bchapter}
\bauthor{\bsnm{Kulkarni}, \binits{S.G.}},
\bauthor{\bsnm{Liu}, \binits{G.}},
\bauthor{\bsnm{Ramakrishnan}, \binits{K.}},
\bauthor{\bsnm{Arumaithurai}, \binits{M.}},
\bauthor{\bsnm{Wood}, \binits{T.}},
\bauthor{\bsnm{Fu}, \binits{X.}}:
\bctitle{Reinforce: Achieving efficient failure resiliency for network function virtualization based services}.
In: \bbtitle{Proceedings of the 14th International Conference on Emerging Networking Experiments and Technologies},
pp. \bfpage{41}--\blpage{53}
(\byear{2018})
\end{bchapter}
\endbibitem

\bibitem[\protect\citeauthoryear{Bailis and Ghodsi}{2013}]{bailis2013eventual}
\begin{barticle}
\bauthor{\bsnm{Bailis}, \binits{P.}},
\bauthor{\bsnm{Ghodsi}, \binits{A.}}:
\batitle{Eventual consistency today: Limitations, extensions, and beyond}.
\bjtitle{Communications of the ACM}
\bvolume{56}(\bissue{5}),
\bfpage{55}--\blpage{63}
(\byear{2013})
\end{barticle}
\endbibitem

\bibitem[\protect\citeauthoryear{Burckhardt}{2014}]{burckhardt2014principles}
\begin{barticle}
\bauthor{\bsnm{Burckhardt}, \binits{S.}}:
\batitle{Principles of eventual consistency}.
\bjtitle{Foundations and Trends{\textregistered} in Programming Languages}
\bvolume{1}(\bissue{1-2}),
\bfpage{1}--\blpage{150}
(\byear{2014})
\end{barticle}
\endbibitem

\bibitem[\protect\citeauthoryear{Vogels}{2008}]{vogels2008eventually}
\begin{barticle}
\bauthor{\bsnm{Vogels}, \binits{W.}}:
\batitle{Eventually consistent: Building reliable distributed systems at a worldwide scale demands trade-offs? between consistency and availability.}
\bjtitle{Queue}
\bvolume{6}(\bissue{6}),
\bfpage{14}--\blpage{19}
(\byear{2008})
\end{barticle}
\endbibitem

\bibitem[\protect\citeauthoryear{Vukolic}{2016}]{vukolic2016eventually}
\begin{barticle}
\bauthor{\bsnm{Vukolic}, \binits{M.}}:
\batitle{Eventually returning to strong consistency}.
\bjtitle{IEEE Data Eng. Bull.}
\bvolume{39}(\bissue{1}),
\bfpage{39}--\blpage{44}
(\byear{2016})
\end{barticle}
\endbibitem

\bibitem[\protect\citeauthoryear{Adve and Gharachorloo}{1996}]{adve1996shared}
\begin{barticle}
\bauthor{\bsnm{Adve}, \binits{S.V.}},
\bauthor{\bsnm{Gharachorloo}, \binits{K.}}:
\batitle{Shared memory consistency models: A tutorial}.
\bjtitle{Computer}
\bvolume{29}(\bissue{12}),
\bfpage{66}--\blpage{76}
(\byear{1996})
\end{barticle}
\endbibitem

\bibitem[\protect\citeauthoryear{Mosberger}{1993}]{mosberger1993memory}
\begin{barticle}
\bauthor{\bsnm{Mosberger}, \binits{D.}}:
\batitle{Memory consistency models}.
\bjtitle{ACM SIGOPS Operating Systems Review}
\bvolume{27}(\bissue{1}),
\bfpage{18}--\blpage{26}
(\byear{1993})
\end{barticle}
\endbibitem

\bibitem[\protect\citeauthoryear{Dixit et~al.}{2013}]{dixit2013towards}
\begin{barticle}
\bauthor{\bsnm{Dixit}, \binits{A.}},
\bauthor{\bsnm{Hao}, \binits{F.}},
\bauthor{\bsnm{Mukherjee}, \binits{S.}},
\bauthor{\bsnm{Lakshman}, \binits{T.}},
\bauthor{\bsnm{Kompella}, \binits{R.}}:
\batitle{Towards an elastic distributed sdn controller}.
\bjtitle{ACM SIGCOMM Computer Communication Review}
\bvolume{43}(\bissue{4}),
\bfpage{7}--\blpage{12}
(\byear{2013})
\end{barticle}
\endbibitem

\bibitem[\protect\citeauthoryear{Carpenter and Brim}{2002}]{carpenter2002middleboxes}
\begin{botherref}
\oauthor{\bsnm{Carpenter}, \binits{B.}},
\oauthor{\bsnm{Brim}, \binits{S.}}:
Middleboxes: Taxonomy and issues.
Technical report,
RFC 3234, February
(2002)
\end{botherref}
\endbibitem

\bibitem[\protect\citeauthoryear{Sherry et~al.}{2012}]{sherry2012making}
\begin{barticle}
\bauthor{\bsnm{Sherry}, \binits{J.}},
\bauthor{\bsnm{Hasan}, \binits{S.}},
\bauthor{\bsnm{Scott}, \binits{C.}},
\bauthor{\bsnm{Krishnamurthy}, \binits{A.}},
\bauthor{\bsnm{Ratnasamy}, \binits{S.}},
\bauthor{\bsnm{Sekar}, \binits{V.}}:
\batitle{Making middleboxes someone else's problem: Network processing as a cloud service}.
\bjtitle{ACM SIGCOMM Computer Communication Review}
\bvolume{42}(\bissue{4}),
\bfpage{13}--\blpage{24}
(\byear{2012})
\end{barticle}
\endbibitem

\bibitem[\protect\citeauthoryear{Szalay et~al.}{2019}]{szalay2019industrial}
\begin{bchapter}
\bauthor{\bsnm{Szalay}, \binits{M.}},
\bauthor{\bsnm{Nagy}, \binits{M.}},
\bauthor{\bsnm{G{\'e}hberger}, \binits{D.}},
\bauthor{\bsnm{Kiss}, \binits{Z.}},
\bauthor{\bsnm{M{\'a}tray}, \binits{P.}},
\bauthor{\bsnm{N{\'e}meth}, \binits{F.}},
\bauthor{\bsnm{Pongr{\'a}cz}, \binits{G.}},
\bauthor{\bsnm{R{\'e}tv{\'a}ri}, \binits{G.}},
\bauthor{\bsnm{Toka}, \binits{L.}}:
\bctitle{Industrial-scale stateless network functions}.
In: \bbtitle{Proceedings of the 2019 IEEE 12th International Conference on Cloud Computing (CLOUD)},
pp. \bfpage{383}--\blpage{390}
(\byear{2019}).
\bcomment{IEEE}
\end{bchapter}
\endbibitem

\bibitem[\protect\citeauthoryear{Stoica et~al.}{2003}]{stoica2003chord}
\begin{barticle}
\bauthor{\bsnm{Stoica}, \binits{I.}},
\bauthor{\bsnm{Morris}, \binits{R.}},
\bauthor{\bsnm{Liben-Nowell}, \binits{D.}},
\bauthor{\bsnm{Karger}, \binits{D.R.}},
\bauthor{\bsnm{Kaashoek}, \binits{M.F.}},
\bauthor{\bsnm{Dabek}, \binits{F.}},
\bauthor{\bsnm{Balakrishnan}, \binits{H.}}:
\batitle{Chord: a scalable peer-to-peer lookup protocol for internet applications}.
\bjtitle{IEEE/ACM Transactions on networking}
\bvolume{11}(\bissue{1}),
\bfpage{17}--\blpage{32}
(\byear{2003})
\end{barticle}
\endbibitem

\bibitem[\protect\citeauthoryear{S.}{1995}]{plank1994libckpt}
\begin{botherref}
\oauthor{\bsnm{S.}, \binits{P.J.}}:
Libckpt : Transparent checkpointing under unix.
Proceedings of the Usenix Winter Technical Conference,
213--223
(1995)
\end{botherref}
\endbibitem

\bibitem[\protect\citeauthoryear{Cachin et~al.}{2016}]{cachin2016non}
\begin{botherref}
\oauthor{\bsnm{Cachin}, \binits{C.}},
\oauthor{\bsnm{Schubert}, \binits{S.}},
\oauthor{\bsnm{Vukoli{\'c}}, \binits{M.}}:
Non-determinism in byzantine fault-tolerant replication.
arXiv preprint arXiv:1603.07351
(2016)
\end{botherref}
\endbibitem

\bibitem[\protect\citeauthoryear{Samaras et~al.}{1995}]{2pc}
\begin{barticle}
\bauthor{\bsnm{Samaras}, \binits{G.}},
\bauthor{\bsnm{Britton}, \binits{K.}},
\bauthor{\bsnm{Citron}, \binits{A.}},
\bauthor{\bsnm{Mohan}, \binits{C.}}:
\batitle{Two-phase commit optimizations in a commercial distributed environment}.
\bjtitle{Distributed and Parallel Databases}
\bvolume{3}(\bissue{4}),
\bfpage{325}--\blpage{360}
(\byear{1995})
\end{barticle}
\endbibitem

\bibitem[\protect\citeauthoryear{Gill et~al.}{2011}]{gill2011understanding}
\begin{bchapter}
\bauthor{\bsnm{Gill}, \binits{P.}},
\bauthor{\bsnm{Jain}, \binits{N.}},
\bauthor{\bsnm{Nagappan}, \binits{N.}}:
\bctitle{Understanding network failures in data centers: measurement, analysis, and implications}.
In: \bbtitle{Proceedings of the ACM SIGCOMM 2011 Conference},
pp. \bfpage{350}--\blpage{361}
(\byear{2011})
\end{bchapter}
\endbibitem

\bibitem[\protect\citeauthoryear{Potharaju and Jain}{2013}]{potharaju2013demystifying}
\begin{bchapter}
\bauthor{\bsnm{Potharaju}, \binits{R.}},
\bauthor{\bsnm{Jain}, \binits{N.}}:
\bctitle{Demystifying the dark side of the middle: a field study of middlebox failures in datacenters}.
In: \bbtitle{Proceedings of the 2013 Conference on Internet Measurement Conference},
pp. \bfpage{9}--\blpage{22}
(\byear{2013})
\end{bchapter}
\endbibitem

\bibitem[\protect\citeauthoryear{Gunawi et~al.}{2014}]{gunawi2014bugs}
\begin{bchapter}
\bauthor{\bsnm{Gunawi}, \binits{H.S.}},
\bauthor{\bsnm{Hao}, \binits{M.}},
\bauthor{\bsnm{Leesatapornwongsa}, \binits{T.}},
\bauthor{\bsnm{Patana-anake}, \binits{T.}},
\bauthor{\bsnm{Do}, \binits{T.}},
\bauthor{\bsnm{Adityatama}, \binits{J.}},
\bauthor{\bsnm{Eliazar}, \binits{K.J.}},
\bauthor{\bsnm{Laksono}, \binits{A.}},
\bauthor{\bsnm{Lukman}, \binits{J.F.}},
\bauthor{\bsnm{Martin}, \binits{V.}},
\bauthor{\bsnm{Satria}, \binits{A.D.}}:
\bctitle{What bugs live in the cloud? a study of 3000+ issues in cloud systems}.
In: \bbtitle{Proceedings of the ACM Symposium on Cloud Computing},
pp. \bfpage{1}--\blpage{14}
(\byear{2014})
\end{bchapter}
\endbibitem

\bibitem[\protect\citeauthoryear{Gunawi et~al.}{2016}]{gunawi2016does}
\begin{bchapter}
\bauthor{\bsnm{Gunawi}, \binits{H.S.}},
\bauthor{\bsnm{Hao}, \binits{M.}},
\bauthor{\bsnm{Suminto}, \binits{R.O.}},
\bauthor{\bsnm{Laksono}, \binits{A.}},
\bauthor{\bsnm{Satria}, \binits{A.D.}},
\bauthor{\bsnm{Adityatama}, \binits{J.}},
\bauthor{\bsnm{Eliazar}, \binits{K.J.}}:
\bctitle{Why does the cloud stop computing? lessons from hundreds of service outages}.
In: \bbtitle{Proceedings of the 7th ACM Symposium on Cloud Computing},
pp. \bfpage{1}--\blpage{16}
(\byear{2016})
\end{bchapter}
\endbibitem

\bibitem[\protect\citeauthoryear{Sahoo et~al.}{2010}]{sahoo2010empirical}
\begin{bchapter}
\bauthor{\bsnm{Sahoo}, \binits{S.K.}},
\bauthor{\bsnm{Criswell}, \binits{J.}},
\bauthor{\bsnm{Adve}, \binits{V.}}:
\bctitle{An empirical study of reported bugs in server software with implications for automated bug diagnosis}.
In: \bbtitle{Proceedings of the 2010 ACM/IEEE 32nd International Conference on Software Engineering},
vol. \bseriesno{1},
pp. \bfpage{485}--\blpage{494}
(\byear{2010}).
\bcomment{IEEE}
\end{bchapter}
\endbibitem

\bibitem[\protect\citeauthoryear{}{}]{pyqueue}
\begin{botherref}
Python's Queue Module, a Synchronized Queue Class.
\url{https://docs.python.org/3/library/queue.html}
\end{botherref}
\endbibitem

\bibitem[\protect\citeauthoryear{}{}]{hazeclast}
\begin{botherref}
Distributed Computation and Storage Platform.
\url{https://github.com/hazelcast/hazelcast}
\end{botherref}
\endbibitem

\bibitem[\protect\citeauthoryear{}{}]{packet-sender}
\begin{botherref}
Packet Sender, an Open Source Utility to Allow Sending and Receiving TCP and UDP Packets.
\url{https://packetsender.com/}
\end{botherref}
\endbibitem

\bibitem[\protect\citeauthoryear{}{}]{twisted}
\begin{botherref}
Twisted, An Event-driven Networking Engine.
\url{https://twisted.org/}
\end{botherref}
\endbibitem

\bibitem[\protect\citeauthoryear{Paxson}{1999}]{bro}
\begin{barticle}
\bauthor{\bsnm{Paxson}, \binits{V.}}:
\batitle{Bro: a system for detecting network intruders in real-time}.
\bjtitle{Computer networks}
\bvolume{31}(\bissue{23-24}),
\bfpage{2435}--\blpage{2463}
(\byear{1999})
\end{barticle}
\endbibitem

\bibitem[\protect\citeauthoryear{}{}]{prads}
\begin{botherref}
Passive Real-time Asset Detection System.
\url{https://github.com/gamelinux/prads}
\end{botherref}
\endbibitem

\end{thebibliography}

\end{document}